\documentclass[aps,prd,superscriptaddress,nofootinbib,preprint]{revtex4}

\usepackage{ulem}
\usepackage{color}
\usepackage{graphicx}
\usepackage{multirow}
\usepackage{hyperref}

\usepackage{amsmath,amssymb,amsthm,amsxtra,overpic,bbm,bm,float,epsfig}

\begin{document}

\title{A Model-Independent Approach to the Reconstruction of Multi-flavor Supernova Neutrino Energy Spectra}

\author{Hui-Ling Li}
\email{lihuiling@ihep.ac.cn}
\affiliation{Institute of High Energy Physics, Chinese Academy of Sciences, Beijing 100049, China}

\author{Xin Huang}
\email{huangxin@ihep.ac.cn}
\affiliation{Institute of High Energy Physics, Chinese Academy of Sciences, Beijing 100049, China}
\affiliation{School of Physical Sciences, University of Chinese Academy of Sciences, Beijing 100049, China}

\author{Yu-Feng Li}
\email{liyufeng@ihep.ac.cn}
\affiliation{Institute of High Energy Physics, Chinese Academy of Sciences, Beijing 100049, China}
\affiliation{School of Physical Sciences, University of Chinese Academy of Sciences, Beijing 100049, China}

\author{Liang-Jian Wen}
\email{wenlj@ihep.ac.cn}
\affiliation{Institute of High Energy Physics, Chinese Academy of Sciences, Beijing 100049, China}

\author{Shun Zhou}
\email{zhoush@ihep.ac.cn}
\affiliation{Institute of High Energy Physics, Chinese Academy of Sciences, Beijing 100049, China}
\affiliation{School of Physical Sciences, University of Chinese Academy of Sciences, Beijing 100049, China}

\begin{abstract}
The model-independent reconstruction of the energy spectra of $\overline{\nu}^{}_e$, $\nu^{}_e$ and $\nu^{}_x$ (i.e., $\nu^{}_\mu$, $\nu^{}_\tau$ and their antiparticles) from the future observation of a galactic core-collapse supernova (SN) is of crucial importance to understand the microscopic physics of SN explosions. To this end, we propose a practically useful method to combine the multi-channel detection of SN neutrinos in a large liquid-scintillator detector (e.g., JUNO), namely, the inverse beta decay $\overline{\nu}^{}_e + p \to e^+ + n$, the elastic neutrino-proton scattering $\nu + p \to \nu + p$ and the elastic neutrino-electron scattering $\nu + e^- \to \nu + e^-$, and reconstruct the energy spectra of $\overline{\nu}^{}_e$, $\nu^{}_e$ and $\nu^{}_x$ by making the best use of the observational data in those three channels. In addition, the neutrino energy spectra from the numerical simulations of the delayed neutrino-driven SN explosions are implemented to demonstrate the robustness of our method. Taking the ordinary matter effects into account, we also show how to extract the initial neutrino energy spectra in the presence of neutrino flavor conversions.
\end{abstract}

\maketitle
\section{INTRODUCTION}
Two dozens of neutrino events from Supernova (SN) 1987A, as observed by Kamiokande-II~\cite{Hirata1987}, Irvine-Michigan-Brookhaven~\cite{Bionta1987} and Baskan~\cite{Alekseev1988}, have essentially confirmed the basic idea of the delayed neutrino-driven explosion mechanism for core-collapse SNe~\cite{Colgate1966, Bethe1990, Woosley2002, Janka2006, Janka2017}. All six flavors of neutrinos and antineutrinos are emitted from the SN core, carrying away most of the gravitational binding energy and bringing ample information about their production and other microscopic physics~\cite{Keil:2002in,Buras:2002wt,Janka:2017vlw}. When propagating outward, SN neutrinos could experience collective flavor conversions caused by the coherent neutrino-neutrino scattering that can take place in the SN environment with an extremely-high neutrino density~\cite{Pantaleone:1992eq, Samuel:1993uw, Duan:2005cp, Duan:2006an, Hannestad:2006nj, Raffelt:2007yz, Duan:2009cd, Duan:2010bg, Chakraborty:2016yeg}, besides the ordinary Mikheyev-Smirnov-Wolfenstein (MSW) matter effects~\cite{Wolfenstein:1977ue, Mikheev:1986gs}.

To diagnose the true pattern of neutrino flavor conversions and even pin down the SN explosion mechanism~\cite{Mirizzi:2015eza}, we have to detect SN neutrinos of all flavors and fully reconstruct their energy spectra. However, the sparse data from SN 1987A do not suffice for such a purpose~\cite{Vissani2014}. For a galactic core-collapse SN at a distance around 10~kpc~\cite{Adams2013}, a few large water-Cherenkov (WC), liquid-scintillator (LS) and liquid-argon time projection chamber (LAr-TPC) neutrino detectors worldwide that are in operation (e.g., Super-Kamiokande~\cite{Okumura2016}, Borexino~\cite{DAngelo2016} and KamLAND~\cite{Tolich2011} ) or under construction (e.g., Hyper-Kamiokande~\cite{Yokoyama2017}, JUNO~\cite{An:2015jdp}, and DUNE~\cite{Acciarri2015}) will register a large number of neutrino events and have a great potential to provide complete flavor information on SN neutrinos. First of all, the energy spectrum of $\overline{\nu}^{}_{e}$ can be unambiguously determined via the inverse beta decay channel $\overline{\nu}^{}_{e}+p\rightarrow e^{+}+n$ (IBD) in both WC and LS detectors. Then, the $\nu^{}_{e}$ energy spectrum can be well measured in the LAr-TPC through the charge-current interaction $\nu^{}_e + ~^{40}{\rm Ar} \to e^- + ~^{40}{\rm K}^*$, and in the WC detector via the elastic neutrino-electron scattering $\nu + e^- \to \nu + e^-$ ($e$ES), which receives the contributions from all neutrino flavors but is most sensitive to $\nu^{}_e$ because of the largest cross section. Finally, the energy spectrum of $\nu^{}_{x}$, which collectively denotes $\nu^{}_{\mu}$, $\nu^{}_{\tau}$ and their antiparticles, can be partially extracted from the elastic neutrino-proton scattering channel $\nu + p \rightarrow \nu + p$ ($p$ES) in the LS detectors, as first proposed in Ref.~\cite{Beacom:2002hs} and further studied in Ref.~\cite{Dasgupta:2011wg}.

In the previous work~\cite{Huiling2018}, we have already shown that it is possible to accomplish a complete reconstruction of the energy spectra of SN neutrinos $\overline{\nu}^{}_{e}$, $\nu^{}_{e}$ and $\nu^{}_{x}$ in a single large LS detector. One salient feature of the LS detector is its low energy threshold, which renders it capable of observing the recoiled protons resulting from the $p$ES process that is mainly sensitive to SN $\nu^{}_x$. Therefore, the energy spectra of $\overline{\nu}^{}_{e}$, $\nu^{}_{e}$ and $\nu^{}_{x}$ can essentially be extracted from the IBD, $e$ES and $p$ES events, respectively. In this paper, we have improved the previous study in Ref.~\cite{Huiling2018} in several important aspects. First, while the SN $\overline{\nu}^{}_e$ energy spectrum can be precisely determined via the IBD data, the energy spectrum of $\nu^{}_e$ cannot be accurately extracted from the $e$ES data. The key point is that although the $e$ES cross section of $\nu^{}_e$ is about six times larger than that of $\nu^{}_\mu$ or $\nu^{}_\tau$ (or their antiparticles), the total contribution to the $e$ES events from the latter four neutrino flavors is obviously significant. In Ref.~\cite{Huiling2018}, it has been assumed that $\nu^{}_e$ dominates over all other flavors in the $e$ES channel and the detector response matrix between the initial neutrino energy and the observed event energy is approximately taken to be universal for all neutrinos. In the present work, this assumption is relaxed and the full detector response matrix will be implemented. Second, the flavor conversions of SN neutrinos, although only the ordinary MSW matter effects in the SN mantle are taken into account, serve as another complication for the spectrum unfolding of multi-flavor neutrinos. Third, the central idea of our reconstruction method is to treat $\overline{\nu}^{}_e$, $\nu^{}_e$ and $\nu^{}_x$ energy spectra on the same footing in all three reaction channels, and build the overall detector response matrix according to their individual interactions with the target particles in the LS. We stress that such a model-independent  approach is also applicable to solar neutrinos (with two flavors $\nu^{}_e$ and $\nu^{}_{\mu/\tau}$) and ultrahigh-energy cosmic neutrinos (with three flavors $\nu^{}_e/\overline{\nu}^{}_e$, $\nu^{}_\mu/\overline{\nu}^{}_\mu$ and $\nu^{}_\tau/\overline{\nu}^{}_\tau$), when the statistics is sufficiently large in the relevant next-generation experiments.

The remaining part of our paper is organized as follows. After a brief description of SN neutrino detection in the LS detector in Sec. II, we introduce our reconstruction method in Sec. III. Then, we present in Sec. IV the unfolding results of SN neutrino energy spectra, and investigate the impact of the energy threshold of the detector and the dependence on the numerical models of SN neutrinos. Moreover, the flavor conversions of SN neutrinos in the presence of MSW matter effects are discussed. Finally, we summarize our main results and conclude in Sec. V.

\section{SUPERNOVA NEUTRINO EVENTS}

In a core-collapse SN, the gravitational potential energy about $3\times 10^{53}$ erg is released and $99\%$ of it is carried away by neutrinos. Neutrinos of all flavors with energies of a few tens MeV are emitted. The duration of neutrino emission lasts for about ten seconds in three distinct phases, namely, the early-time neutronization burst, the accretion phase and the cooling phase. In this work, however, we focus on the time-integrated SN neutrino energy spectra and their reconstruction from simulated experimental data.

In LS detectors, the relevant reactions for SN neutrinos have been studied in detail in Refs.~\cite{An:2015jdp, Lu:2016ipr, Lujan-Peschard:2014lta}. Although the charged- and neutral-current interactions with the $^{12}{\rm C}$ and $^{13}{\rm C}$ nuclei are also available therein, the corresponding numbers of neutrino events are sub-dominant~\cite{Lu:2016ipr}. For simplicity, we consider only three dominant channels, i.e., the IBD, $e$ES, and $p$ES. The neutrino event rates in these channels for a JUNO-like detector have been calculated in Ref.~\cite{Huiling2018} and will be recapped below for completeness.

\subsection{SN Neutrino Spectra}

The differential fluences or time-integrated energy spectra of SN neutrinos can reasonably be described by the Keil-Raffelt-Janka (KRJ) parametrization~\cite{Keil:2002in}
\begin{eqnarray}
\frac{{\rm d}F^{}_\alpha}{{\rm d}E^{}_\alpha} = \frac{3.5\times 10^{13}}{{\rm cm}^{2}~{\rm MeV}} \cdot \frac{1}{4\pi D^2} \frac{\varepsilon^{}_\alpha}{\langle E^{}_\alpha \rangle} \frac{E^{\gamma^{}_\alpha}_\alpha}{\Gamma(1+\gamma^{}_\alpha)} \left(\frac{1+\gamma^{}_\alpha}{\langle E^{}_\alpha \rangle}\right)^{1+\gamma^{}_\alpha} \exp\left[ - (1+\gamma^{}_\alpha) \frac{E^{}_\alpha}{\langle E^{}_\alpha \rangle}\right] \; ,
\label{eq:KRS}
\end{eqnarray}
where the subscript $\alpha$ runs over three neutrino flavors $\nu^{}_e$, $\overline{\nu}^{}_e$ and $\nu^{}_x$, and $\varepsilon^{}_\alpha$ is the total neutrino energy for each flavor in units of $5\times 10^{52}~{\rm erg}$. In addition, the distance $D$ to a galactic SN is normalized to a typical value of $10~{\rm kpc}$, the neutrino energy $E^{}_\alpha$ and average energy $\langle E^{}_\alpha \rangle$ are measured in MeV. As an example for the analytical model of SN neutrino energy spectra, the spectral index $\gamma^{}_\alpha = 3$ will always be adopted. In the assumption of energy equipartition, the total gravitational binding energy $E^{}_{\rm g} = 3\times 10^{53}~{\rm erg}$ released in the SN explosion within ten seconds is shared by $\nu^{}_e$, $\nu^{}_\mu$, $\nu^{}_\tau$ and their antiparticles, i.e., $\varepsilon^{}_\alpha \approx 5\times 10^{52}~{\rm erg}$. Whenever the analytical model is referred to, the nominal values of neutrino average energies will be set to $\langle E^{}_{\nu^{}_e}\rangle = 12~{\rm MeV}$, $\langle E^{}_{\overline{\nu}^{}_e} \rangle = 14~{\rm MeV}$ and $\langle E^{}_{\nu^{}_x} \rangle = 16~{\rm MeV}$.

For numerical models of SN neutrino spectra, we make use of the simulation results from the Japan group~\cite{Nakazato:2013}, where the time and energy distributions of all neutrino flavors are provided for a wide range of progenitor star masses and different values of metallicity. In this case, we numerically integrate the two-dimensional distribution over time to obtain the neutrino energy spectra for each simulation model.

It is worthwhile to mention that all the neutrino energy spectra from either the analytical model with the KRJ parametrization or the numerical models from the simulations of SN explosions are implemented as the original ones in the following discussions. Only in Sec. IV C do we investigate the impact of neutrino flavor conversions on the reconstruction strategy.

\subsection{SN Neutrino Events}

In Ref~\cite{Huiling2018}, the SN neutrino events of IBD, $p$ES and $e$ES channels have been simulated for the JUNO-like detector with a fiducial mass of 20 kiloton, of which 12$\%$ are protons and 88$\%$ are carbon nuclei. The detector energy resolution is taken to be $3\%/\sqrt{E^{}_{\rm o}/({\rm MeV})}$ with $E_{\rm_{o}}$ being the observed energy. Compared to the WC and LAr-TPC detectors, the LS detectors are able to reach a much lower energy threshold. In this work, an energy threshold of 0.2 MeV will be adopted as default in our toy Monte Carlo (MC) simulations. The quenching effect for protons in LS is carefully taken into account. Although the recoil energy of protons is severely quenched in the LS, one can measure it down to 0.2 MeV through a good control of radioactive backgrounds as well as dark noises of photomultiplier tubes~\cite{An:2015jdp}. For a galactic SN at $D = 10~{\rm kpc}$, such a JUNO-like detector will register about 5000 IBD, 1500 $p$ES and approximately 400 $e$ES events.

The IBD channel is the golden channel for SN neutrino detection in LS detectors due to the time coincidence of the prompt and slow signals and the large cross section of the IBD reaction~\cite{Vogel:1999zy, Strumia:2003zx}. This channel is solely sensitive to $\overline{\nu}^{}_{e}$. The other two elastic scattering channels, $p$ES and $e$ES, receive contributions from neutrinos of all flavors $\nu^{}_{e}$, $\overline{\nu}^{}_{e}$ and $\nu^{}_{x}$. For SN neutrinos with a few tens of MeV, the cross section for $p$ES is almost the same for all flavors~\cite{Weinberg:1972tu, Beacom:2002hs}. In the $e$ES channel, the cross sections of electron flavor neutrinos $\nu^{}_{e}$ are larger than that of $\nu_{x}$~\cite{tHooft:1971ucy,Marciano:2003eq} due to the charged-current interaction of the former. The signals of recoiled protons and electrons in the LS detectors, from the $p$ES and $e$ES channel respectively, can be distinguished by utilizing the technique of pulse shape discrimination. In the following discussions, all the events in those three channels are treated ideally without considering the uncertainties from background signals and detection efficiencies. The effects of $^{12}$C-related interaction channels on the SN neutrino spectra reconstruction will be studied in a future separated work.
\begin{figure}[!t]
\begin{center}
\begin{tabular}{l}
\includegraphics[width=0.51\textwidth]{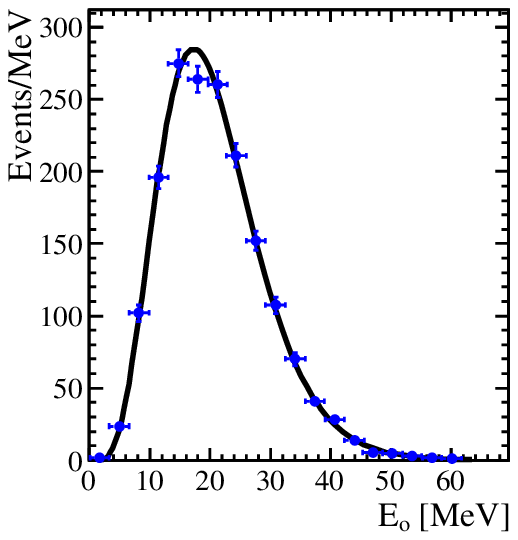}
\hspace{-1cm}
\includegraphics[width=0.51\textwidth]{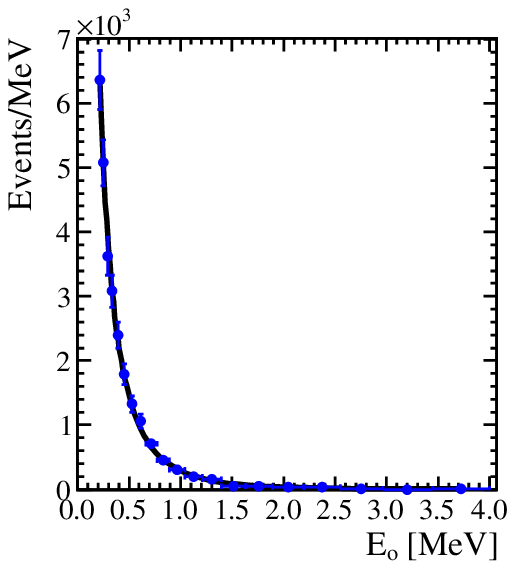}
\\
\includegraphics[width=0.51\textwidth]{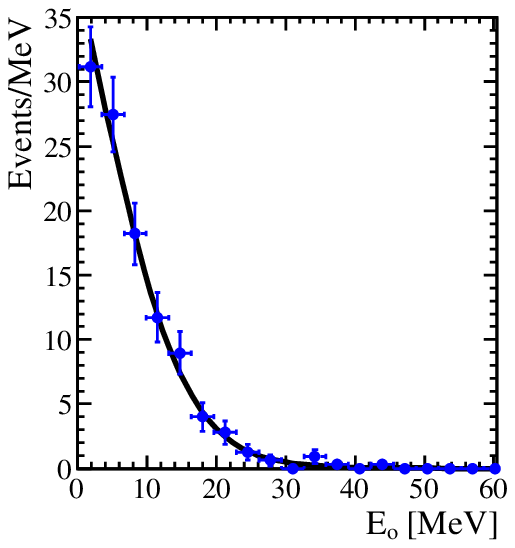}
\hspace{-1cm}
\includegraphics[width=0.51\textwidth]{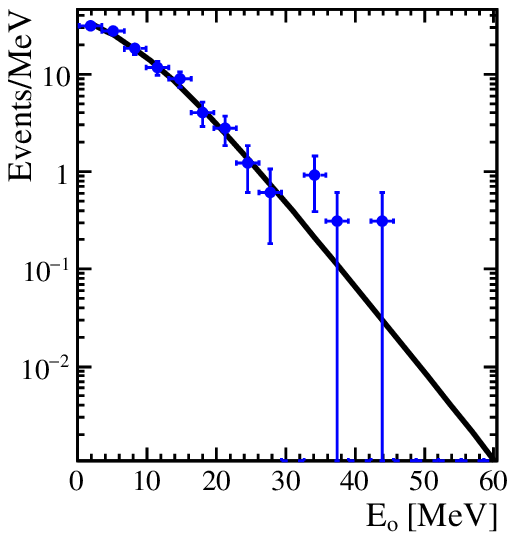}
\end{tabular}
\end{center}
\vspace{-0.5cm}
\caption{The neutrino event spectra at a JUNO-like LS detector for a galactic SN at the distance of 10 kpc, where the first row refers to the IBD (left panel) and $p$ES (right panel) channels while the second row to the $e$ES channel. In the latter case, the event numbers are shown in both linear and logarithmic scales in the left and right panel, respectively.
\label{fig:compDet}}
\end{figure}

For illustration, the IBD and $p$ES event spectra in a JUNO-like LS detector for a galactic SN at a distance of 10~kpc have been shown in the first row of Fig.~\ref{fig:compDet}. In the second row, the $e$ES event spectrum is displayed as well, where the number of events has been given in both linear (left panel) and logarithmic (right panel) scales. In all these plots, the theoretical predictions for the event numbers are denoted by black solid curves, while the toy MC results by blue dots, for which the widths of energy bins and the statistical uncertainties are respectively represented by horizontal and vertical error bars. Note that the toy MC samples are randomly generated within the ROOT framework~\cite{Brun:1997pa} according to the event spectra given in Eqs.~(3), (6) and (8) of Ref.~\cite{Huiling2018}, where an energy resolution of $3\% / \sqrt{E^{}_{\rm o}/({\rm MeV})}$ is always adopted and the analytical SN neutrino flux model with $\langle E^{}_{\nu^{}_e}\rangle = 12~{\rm MeV}$, $\langle E^{}_{\overline{\nu}^{}_e} \rangle = 14~{\rm MeV}$ and $\langle E^{}_{\nu^{}_x} \rangle = 16~{\rm MeV}$ is assumed.

\section{STRATEGY FOR RECONSTRUCTION}

As demonstrated in Ref.~\cite{Huiling2018}, one can reconstruct the SN neutrino spectra of all flavors with three unfolded spectra of IBD, $p$ES and $e$ES channels in one single large LS detector via a simple bin-to-bin separation method. However, as the number of events for $\nu^{}_{e}$ dominates over those for all other flavors in the $e$ES channel, it has been assumed in Ref.~\cite{Huiling2018} that the detector response matrix between the initial neutrino energy and the observed event energy is universal for all neutrinos. In this section, we shall improve the strategy for reconstruction by relaxing this assumption, and extract the energy spectra of all flavor neutrinos directly from the observed spectra of these three channels by using the overall response matrix.

\subsection{The separated analysis}

For comparison, we first summarize the basic idea of Ref.~\cite{Huiling2018} for the spectral reconstruction. For the IBD channel, the observed event spectrum can be calculated as follows
\begin{equation}
N^{}_{\rm{p}} \cdot \bm{D}^{}_{\rm{IBD}} \cdot \bm{\sigma}^{\rm{IBD}}_{\rm{\overline{\nu}^{}_{e}}} \cdot \bm{F}_{\overline{\nu}^{}_{e}} = \bm{S}^{}_{\rm{IBD}} \; ,
\label{eq:IBDmat}
\end{equation}
where $N^{}_{\rm{p}}$ is the number of protons in the LS detector, $\bm{F}^{}_{\overline{\nu}^{}_{e}}$ stands for the spectrum of the SN $\overline{\nu}^{}_{e}$, and $\bm{S}^{}_{\rm{IBD}}$ denotes the observed event spectrum. In addition, $\bm{\sigma}^{\rm{IBD}}_{\overline{\nu}^{}_{e}}$ is the cross section matrix for the IBD reaction, which maps the initial neutrino energy $E^{}_{\nu}$ into the visible energy of the final-state particle $E^{}_{\rm v}$. In Eq.~(\ref{eq:IBDmat}), $\bm{D}_{\rm{IBD}}$ is the probability matrix, which accounts for the detector effects and converts $E^{}_{\rm v}$ into the observed energy $E^{}_{\rm o}$. Note that those matrices and vectors have been constructed by discretizing the relevant continuous energy spectra into finite energy bins. To be explicit, we can recast Eq.~(\ref{eq:IBDmat}) into the matrix form
\begin{equation}
N^{}_{\rm{p}} \cdot
\begin{bmatrix}
D_{11}^{\rm{IBD}} & \dots   & D_{1n_{\rm v}}^{\rm{IBD}}\\
\vdots   & \ddots & \\
D_{n_{\rm o}1}^{\rm{IBD}} &\dots & D_{n_{\rm o}n_{\rm v}}^{\rm{IBD}}
\end{bmatrix} \cdot
\begin{bmatrix}
\sigma^{\overline{\nu}^{}_{e}}_{11} & \dots   & \sigma^{\overline{\nu}^{}_{e}}_{1n_{\nu}}\\
\vdots   & \ddots & \\
\sigma^{\overline{\nu}^{}_{e}}_{n_{\rm v}1} & \dots & \sigma^{\overline{\nu}^{}_{e}}_{n_{\rm v}n_{\nu}}
\end{bmatrix} \cdot
\begin{bmatrix}
F^{\overline{\nu}^{}_{e}}_{11} \\
\vdots  \\
F^{\overline{\nu}^{}_{e}}_{1n_{\nu}}
\end{bmatrix}
=
\begin{bmatrix}
S^{\rm{IBD}}_{11}\\
\vdots\\
S^{\rm{IBD}}_{1n_{\rm o}}
\end{bmatrix} \; ,
\label{eq:IBDMat1}
\end{equation}
where $n^{}_{\rm o}$, $n^{}_{\rm v}$ and $n^{}_{\nu}$ are the number of bins for the observed energy, the visible energy of the final-state particle and the initial energy of SN neutrinos, respectively. Comparing between Eq.~(\ref{eq:IBDmat}) and Eq.~(\ref{eq:IBDMat1}), one can easily identify the definitions of the corresponding matrices.

For practical purposes, we further normalize the cross-section matrix $\bm{\sigma}^{\rm IBD}_{\overline{\nu}_e}$ and the initial spectrum $\bm{F}^{}_{\overline{\nu}_e}$ in the following way
\begin{equation}
\begin{bmatrix}
D_{11}^{\rm{IBD}} & \dots   & D_{1n_{\rm v}}^{\rm{IBD}}\\
\vdots   & \ddots & \\
D_{n_{\rm o}1}^{\rm{IBD}} &\dots & D_{n_{\rm o}n_{\rm v}}^{\rm{IBD}}
\end{bmatrix} \cdot
\begin{bmatrix}
\sigma^{\overline{\nu}^{}_{e}}_{11}/\sigma_{1}& \dots   & \sigma^{\overline{\nu}^{}_{e}}_{1n_{\nu}}/\sigma_{n_{\nu}}\\
\vdots   & \ddots & \\
\sigma^{\overline{\nu}^{}_{e}}_{n_{\rm v}1}/\sigma_{1} & \dots & \sigma^{\overline{\nu}^{}_{e}}_{n_{\rm v}n_{\nu}}/\sigma_{n_{\nu}}
\end{bmatrix} \cdot
\begin{bmatrix}
N_{\rm p}\sigma_{1}F^{\overline{\nu}^{}_{e}}_{11} \\
\vdots  \\
N_{\rm p}\sigma_{n_{\nu}}F^{\overline{\nu}^{}_{e}}_{1n_{\nu}}
\end{bmatrix}
=
\begin{bmatrix}
S^{\rm{IBD}}_{11}\\
\vdots\\
S^{\rm{IBD}}_{1n_{\rm o}}
\end{bmatrix} \; ,
\label{eq:IBDMat2}
\end{equation}
where $\sigma^{}_{i}$ for $i = 1, 2, \cdots, n^{}_\nu$ stands for the total cross section for the incident neutrino with the central energy $E_{\nu}^{i}$ of the $i$-th energy bin. Eq.~(\ref{eq:IBDMat2}) can be rewritten in a more compact form as $\bm{R}^{}_{\rm IBD} \bm{\widehat{F}}_{\overline{\nu}^{}_{e}}^{\rm{IBD}} = \bm{S}^{}_{\rm{IBD}}$ with $\bm{R}^{}_{\rm IBD} \equiv \bm{D}^{}_{\rm{IBD}} \bm{\widehat{\sigma}}_{\rm{IBD}}$, where the definitions of relevant matrices are self-evident. Therefore, the reconstruction of the cross section weighted neutrino spectrum $\bm{\widehat{F}}_{\overline{\nu}^{}_{e}}^{\rm{IBD}}$ from the observed event spectrum $\bm{S}^{}_{\rm{IBD}}$ can be regarded as a linear inverse problem, which can be routinely solved with an unfolding method~\cite{Huiling2018}. Here $\bm{R}_{\rm{IBD}}$ is exactly the detector response matrix in Ref.~\cite{Huiling2018}, which can be built with a large number of simulated events and is independent of SN neutrino models.

As for the $p$ES and $e$ES channels, the observed event spectra receive contributions from neutrinos and antineutrinos of all flavors, and can be calculated in a similar way
\begin{equation}
N^{}_{\rm p(e)} \cdot \bm{D}^{}_{p(e){\rm ES}} \cdot \sum_{\alpha}{\bm{\sigma}^{p(e){\rm ES}}_{\alpha} \bm{F}^{}_{\alpha}} = \bm{S}^{}_{p(e){\rm ES}} \; ,
\label{eq:ESmat}
\end{equation}
where $N^{}_{\rm p(e)}$ denotes the number of protons for $p$ES (electrons for $e$ES), and $\alpha$ refers to different neutrino flavors. In analogy to the IBD channel, we can deal with the $p$ES channel by introducing the detector response matrix $\bm{R}^{}_{p{\rm ES}} \equiv \bm{D}^{}_{p{\rm ES}} \bm{\widehat{\sigma}}^{}_{p{\rm ES}}$ and the cross section weighted spectrum $\sum_\alpha \bm{\widehat{F}}^{p{\rm ES}}_\alpha$. Then, it is straightforward to extract $\sum_\alpha \bm{\widehat{F}}^{p{\rm ES}}_\alpha$ from the observed event spectrum $\bm{R}^{}_{p{\rm ES}} \sum_\alpha \bm{\widehat{F}}^{p{\rm ES}}_\alpha = \bm{S}^{}_{p{\rm ES}}$. Such a treatment is quite reasonable in the $p$ES channel, since neutrinos of all flavors interact with protons via the neutral-current interaction and the corresponding cross sections are almost the same. However, this is obviously not the case for the $e$ES channel, for which the cross section of $\nu^{}_e$-$e^-$ scattering is about two times larger than that of $\overline{\nu}^{}_{e}$-$e^{-}$ scattering and six times larger than that of $\nu^{}_x$-$e^-$ scattering. In Ref.~\cite{Huiling2018}, it has been assumed that $\nu^{}_{e}$ dominates over all other flavors in $e$ES channel and a universal response matrix $\bm{R}^{}_{e\rm{ES}} \equiv \sum^{}_{\alpha} \bm{D}^{}_{e{\rm ES}} \bm{\widehat{\sigma}}^{\alpha}_{e{\rm ES}}$ is then used to achieve the reconstruction of the weighted true spectrum $\sum^{}_\alpha \bm{\widehat{F}}_{\alpha}^{e\rm{ES}}$. This approximation is only valid for the one flavor dominated case. Finally, the energy spectra for different flavor neutrinos can be simply separated bin-by-bin from the unfolded spectra $\bm{\widehat{F}}_{\overline{\nu}^{}_{e}}^{\rm{IBD}}$, $\sum_\alpha \bm{\widehat{F}}_{\alpha}^{p\rm{ES}}$ and $\sum_\alpha \bm{\widehat{F}}_{\alpha}^{e\rm{ES}}$.

Although the cross section of the $e$ES for $\nu^{}_e$ is much larger than that for $\nu^{}_x$ and thus $\nu^{}_e$ gives rise to most of the $e$ES events, the summation of the contributions from four flavors of $\nu^{}_x$ with high energies is not negligible. Due to the quenching effects on recoiled protons, only the high-energy part of the $\nu^{}_x$ spectrum (i.e., above 20 MeV) can be really reconstructed from experimental data. The $e$ES channel will be able to provide useful information about the low-energy part of the $\nu^{}_x$ spectrum. Therefore, it is interesting to have a further look at the reconstruction of neutrino spectra in the $e$ES channel.

\subsection{The combined analysis}

Now we put forward a combined analysis of all three observed spectra from IBD, $p$ES and $e$ES by grouping the multi-flavor neutrino spectra into an overall neutrino spectrum, and likewise for the event spectra. More explicitly, we have
\begin{equation}
\bm{S}^{}_{\rm c} =
\begin{bmatrix}
\bm{S}^{}_{\rm{IBD}} \\  \bm{S}^{}_{p\rm{ES}} \\ \bm{S}^{}_{e\rm{ES}}
\end{bmatrix} \; , \quad
\bm{F}^{}_{\rm c} =
\begin{bmatrix}
\bm{F}^{}_{\nu_{e}} \\  \bm{F}^{}_{\overline{\nu}^{}_{e}} \\  \bm{F}^{}_{\nu_{x}}
\end{bmatrix} \; ,
\end{equation}
where $\bm{S}^{}_{\rm c}$ is the whole event spectrum with $(n^{\rm {IBD}}_{\rm o} + n^{p\rm{ES}}_{\rm o} + n^{e\rm {ES}}_{\rm o})$ bins, and $\bm{F}_{\rm c}$ is the overall neutrino spectrum with $(n^{}_{\nu_{e}} + n^{}_{\overline{\nu}^{}_{e}} + n^{}_{\nu_{x}})$ bins. Therefore, the whole event spectrum $\bm{S}^{}_{\rm c}$, as observed in the LS detector, can be described in one single equation
\begin{equation}
\begin{bmatrix}
 N^{}_{\rm p} \bm{D}^{}_{\rm{IBD}} \bm{\sigma}^{\rm{IBD}}_{\nu_{e}} & N^{}_{\rm p} \bm{D}^{}_{\rm{IBD}} \bm{\sigma}^{\rm{IBD}}_{\overline{\nu}^{}_{e}} & N^{}_{\rm p} \bm{D}^{}_{\rm{IBD}} \sum \bm{\sigma}^{\rm{IBD}}_{\nu_{x}} \\[0.2cm]
 N_{\rm p}\bm{D}_{p\rm{ES}}\bm{\sigma}^{p\rm{ES}}_{\nu_{e}}&N_{\rm p}\bm{D}_{p\rm{ES}}\bm{\sigma}^{p\rm{ES}}_{\overline{\nu}^{}_{e}} & N_{\rm p}\bm{D}_{p\rm{ES}}\sum \bm{\sigma}^{p\rm{ES}}_{\nu_{x}} \\[0.2cm]
N_{\rm e}\bm{D}_{e\rm{ES}}\bm{\sigma}^{e\rm{ES}}_{\nu_{e}} & N_{\rm e}\bm{D}_{e\rm{ES}}\bm{\sigma}^{e\rm{ES}}_{\overline{\nu}^{}_{e}} & N_{\rm e}\bm{D}_{e\rm{ES}} \sum \bm{\sigma}^{e\rm{ES}}_{\nu_{x}}
\end{bmatrix} \cdot
\begin{bmatrix}
\bm{F}_{\nu_{e}} \\[0.3cm] \bm{F}_{\overline{\nu}^{}_{e}} \\[0.3cm] \bm{F}_{\nu_{x}}
\end{bmatrix}
=
\begin{bmatrix}
\bm{S}^{}_{\rm{IBD}} \\[0.3cm] \bm{S}^{}_{p\rm{ES}} \\[0.3cm] \bm{S}^{}_{e\rm{ES}}
\end{bmatrix} \; ,
\label{eq:CombinedMethod}
\end{equation}
where the $(n^{\rm {IBD}}_{\rm o} + n^{p\rm{ES}}_{\rm o} + n^{e\rm {ES}}_{\rm o})\times (n_{\nu_{e}}+n_{\overline{\nu}^{}_{e}}+n_{\nu_{x}})$ matrix on the left-hand side is just the detector response matrix $\bm{R}_{\rm c}$ for the combined analysis, and the sum in its third column is running over $\nu^{}_{\mu}$, $\nu^{}_{\tau}$ and their antiparticles. Some explanations for the structure of $\bm{R}^{}_{\rm c}$ are necessary. First of all, the block matrices $\bm{D}^{}_{\rm IBD} \bm{\sigma}^{\rm IBD}_{\nu_e}$ and ${\bm D}^{}_{\rm IBD} \sum \bm{\sigma}^{\rm IBD}_{\nu_x}$ in the first row of $\bm{R}^{}_{\rm c}$ are actually vanishing, since the IBD reaction takes place only for $\overline{\nu}^{}_e$. Second, for the $p$ES and $e$ES channels, the response matrices have been determined by simulating the interactions of $\nu^{}_e$, $\overline{\nu}^{}_e$ and $\nu^{}_x$ with the target particles in the detector. In particular, in the $e$ES channel, the elastic scattering of $\nu^{}_e$, $\overline{\nu}^{}_e$ and $\nu^{}_x$ with electrons has been investigated individually to produce three matrices in the last row of $\bm{R}^{}_{\rm c}$. Third, it is worthwhile to emphasize that such a combined analysis treats $\nu^{}_{e}$, $\overline{\nu}_{e}$ and $\nu^{}_{x}$ on the same footing and it is independent of SN neutrino models. Moreover, this method can be easily extended to include the observations from the WC and LAr-TPC detectors, leading to a global analysis of all SN neutrino data. This can be realized by adding a new row into $\bm{R}^{}_{\rm c}$, which is determined by the specified interaction channel in a given detector, and accordingly a new row in $\bm{S}^{}_{\rm c}$.

To demonstrate how the combined analysis works, we concentrate on the LS detector. Given Eq.~(\ref{eq:CombinedMethod}), one can immediately apply the spectral unfolding approach to extract $\bm{F}^{}_{\rm c}$. See, e.g., Refs.~\cite{Blobel:1984ku, Zech:2016gca}, for a general review on the unfolding problem in particle physics. In the following, we implement the Singular Value Decomposition (SVD) method with a proper regularization scheme, as proposed in Ref.~\cite{Hocker:1995kb}, to reconstruct SN neutrino spectra. The regularization is important to suppress the spurious oscillating components in the final results. The main strategy for reconstruction is outlined as below:
\begin{figure}
\centering
\includegraphics[scale=0.7]{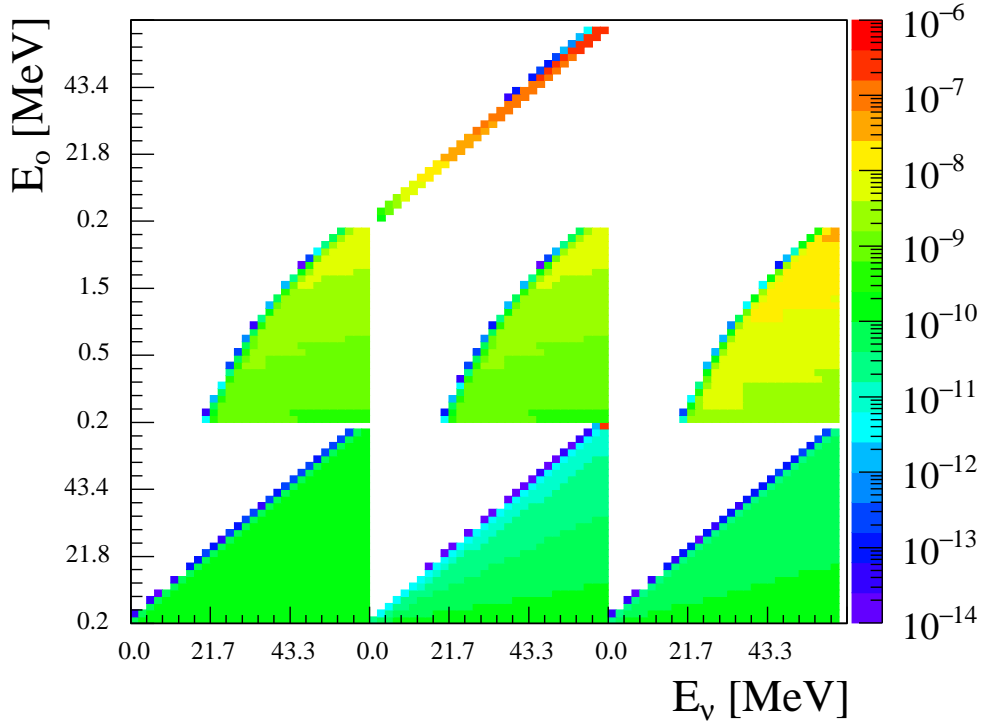}
\vspace{-0.7cm}
\caption{The response matrix for the LS detector, which has been implemented in the combined analysis. Three rows correspond to the IBD (upper), $p$ES (middle) and $e$ES (lower) channels, respectively, while three columns to the $\nu^{}_{e}$ (left), $\overline{\nu}^{}_{e}$ (middle) and 4$\nu^{}_{x}$ (right) spectra. The finite energy resolution of $3\%/\sqrt{E^{}_{\rm o}/({\rm MeV})}$ and the quenching effects in the LS detector have been taken into account, and the energy threshold $E_{\rm o}^{\rm{th}} = 0.2$ MeV has been assumed.}
\label{fig:respMat_noosc}
\end{figure}
\begin{itemize}
\item The detector response matrix for the combined analysis should be constructed, and the results are now depicted in Fig.~\ref{fig:respMat_noosc}. Comparing Fig.~\ref{fig:respMat_noosc} with the matrix $\bm{R}^{}_{\rm c}$ in Eq.~(\ref{eq:CombinedMethod}), one can clearly identify the corresponding block matrices. Obviously, three rows in this figure correspond to the IBD, $p$ES and $e$ES channels, respectively. The energy resolution of the detector and the quenching effects for recoiled protons are taken into account in the simulation. The energy threshold of $E_{\rm o}^{\rm{th}}=0.2$ MeV is assumed, corresponding to the minimal neutrino energies of about 20 MeV in the $p$ES channel and 0.35 MeV in the $e$ES channel. For each one of the nine block matrices in the response matrix, we first fill one histogram with 200 million random events, for which the true neutrino energy spectra have been taken to be flat distributed to avoid any prior information on the input. Then, for the same neutrino energy $E^{}_\nu$, the summation of the events in the histogram in all the bins running through the whole range of the observed energy $E^{}_{\rm o}$, including the overflow and underflow bins, is normalized to one. Afterwards, all the involved bins will be multiplied by the corresponding total cross section $\sigma$ at $E^{}_{\nu}$ and the number of target particles in the specified channel. Finally, all nine histograms generated in this way are shown in Fig.~\ref{fig:respMat_noosc}. Note that the three block matrices for $e$ES channels are different, which demonstrate the limitation of the separated analysis as discussed in the previous subsection.

\item After preparing the response matrix, we then apply the SVD unfolding method to reconstruct of the SN neutrino spectra. In the unfolding process, the binning scheme for the observed event spectrum in each channel depends on the event statistics and should be carefully handled in order to guarantee a comparable number of neutrino events in each bin of the observed energy, as illustrated in Fig.~\ref{fig:compDet} for a SN distance at 10 kpc. For each true neutrino energy spectrum, we employ the equal-size binning scheme but combine the bins at the boundaries due to the limited statistics. The actual binning scheme for the realistic unfolding process can be readily read out from the energy spectra which will be presented in the next section. The practical realization of the SVD unfolding algorithm is based on the TSVDUnfold in ROOT, where a proper regularization parameter is set for this work.
\end{itemize}

As we have mentioned, the response matrix has been built from the simulated neutrino events in each detection channel for a given detector. Therefore, it depends only on the experimental setup and neutrino interactions with the target particles, implying that the combined analysis can be applied to any SN neutrino model, namely, both analytical and numerical ones, and the realistic SN explosion. In the next section, we shall present the final results of the reconstructed SN neutrino spectra.


\section{Reconstruction of Neutrino Spectra}

\begin{figure}[t!]
\vspace{-0.0cm}
\begin{center}
\begin{tabular}{c}
\includegraphics[width=0.47\textwidth,height=0.28\textheight]{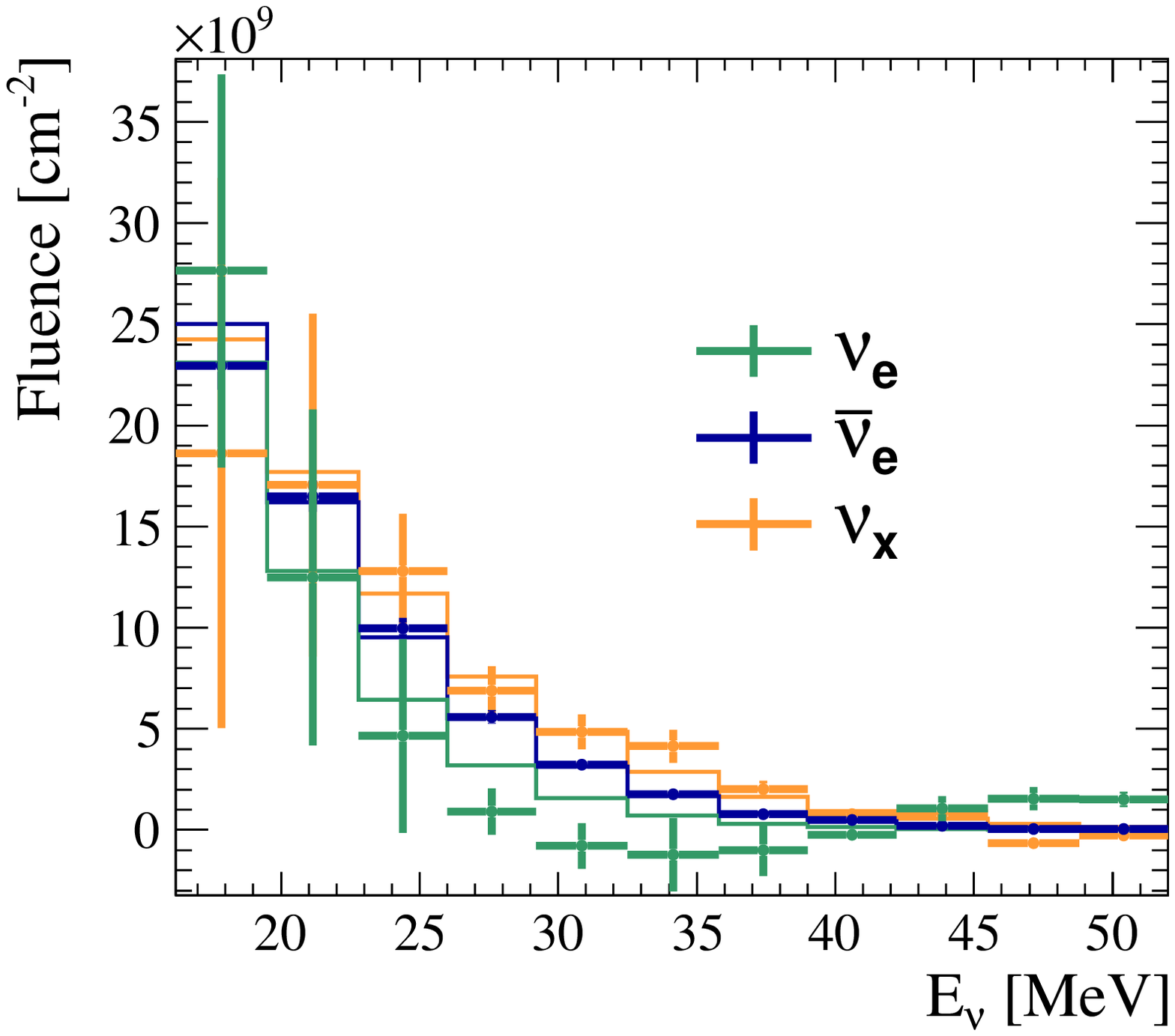}
\includegraphics[width=0.47\textwidth,height=0.28\textheight]{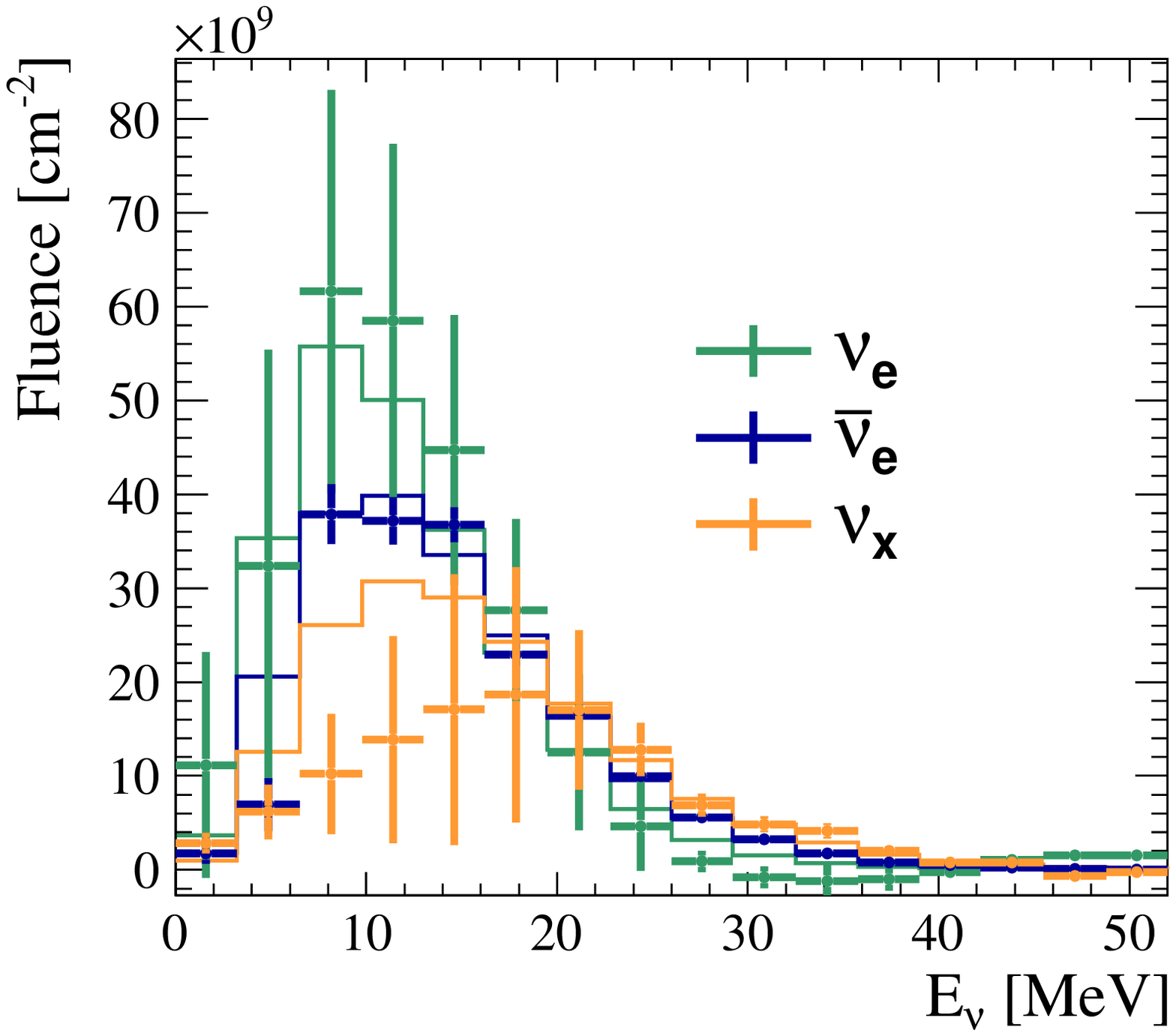}
\\
\includegraphics[width=0.47\textwidth,height=0.28\textheight]{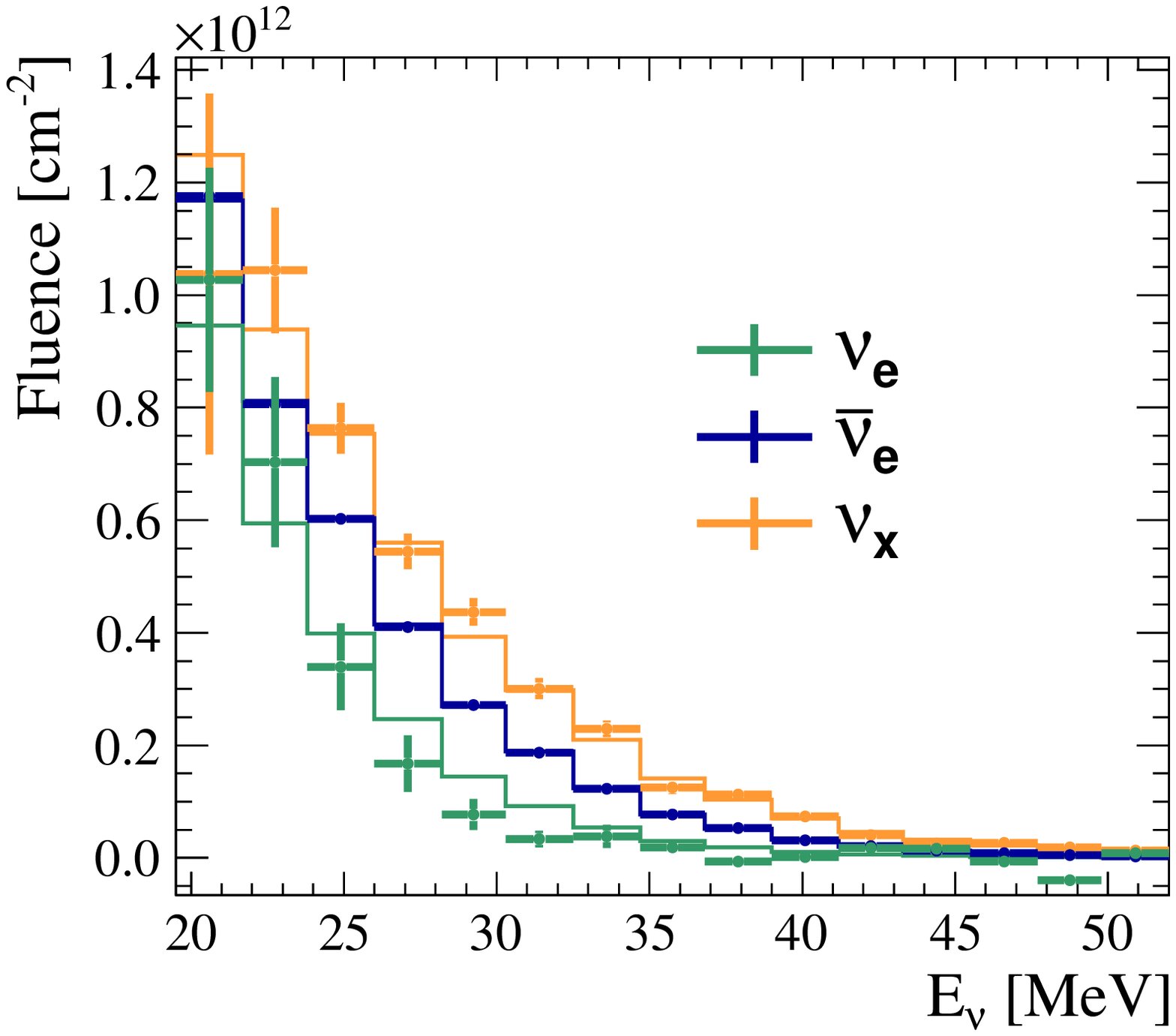}
\includegraphics[width=0.47\textwidth,height=0.28\textheight]{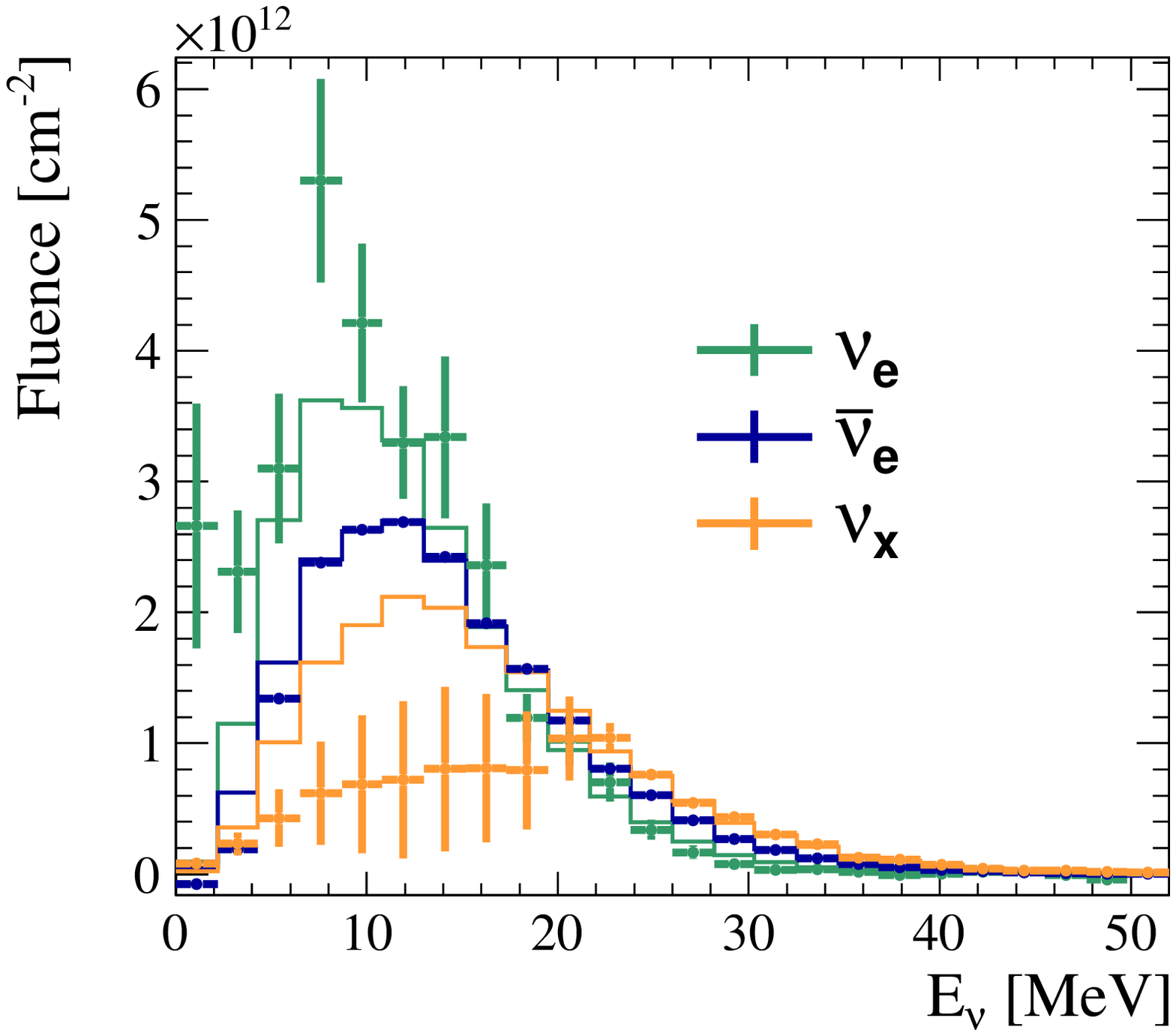}
\\
\includegraphics[width=0.47\textwidth,height=0.28\textheight]{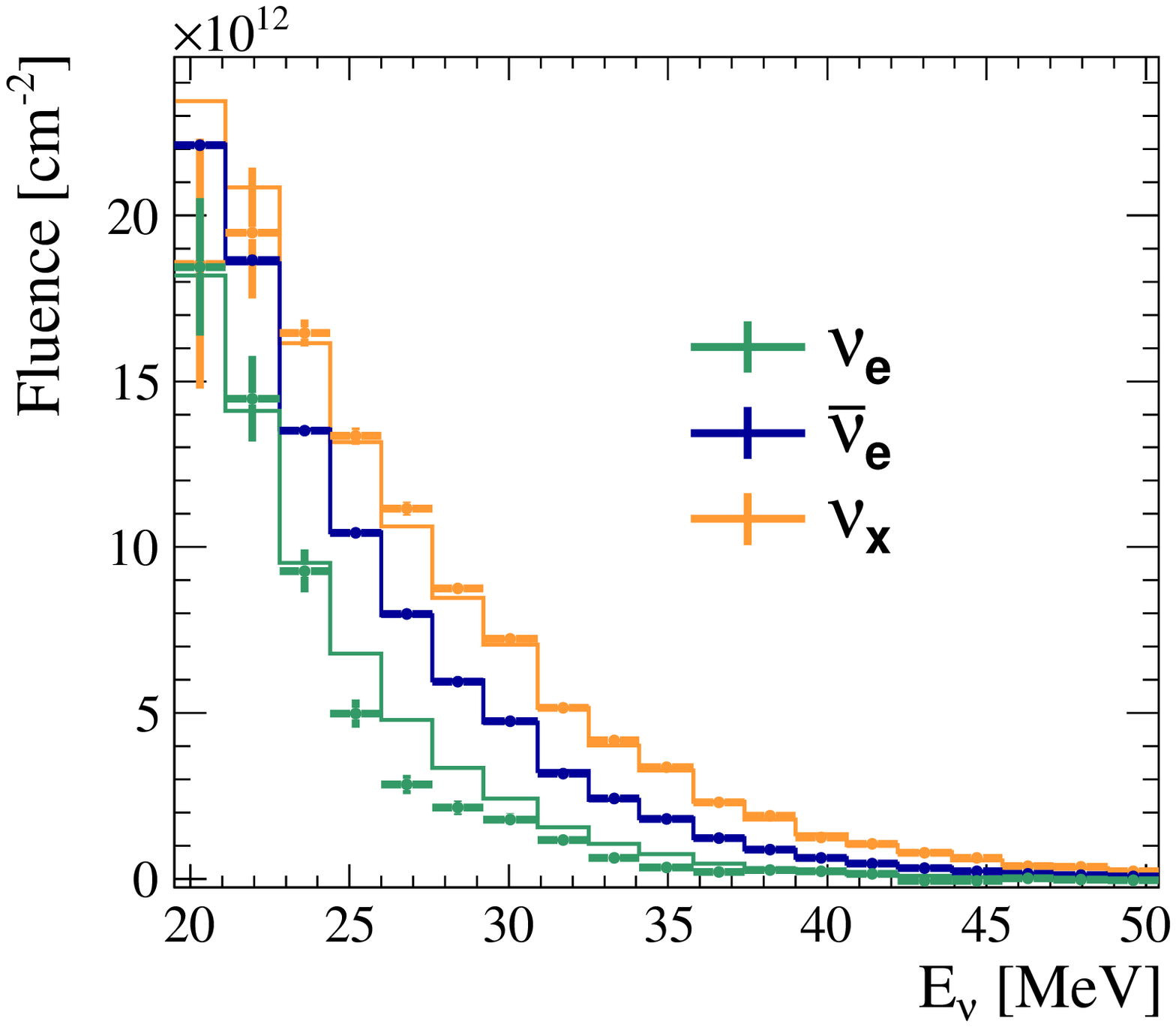}
\includegraphics[width=0.47\textwidth,height=0.28\textheight]{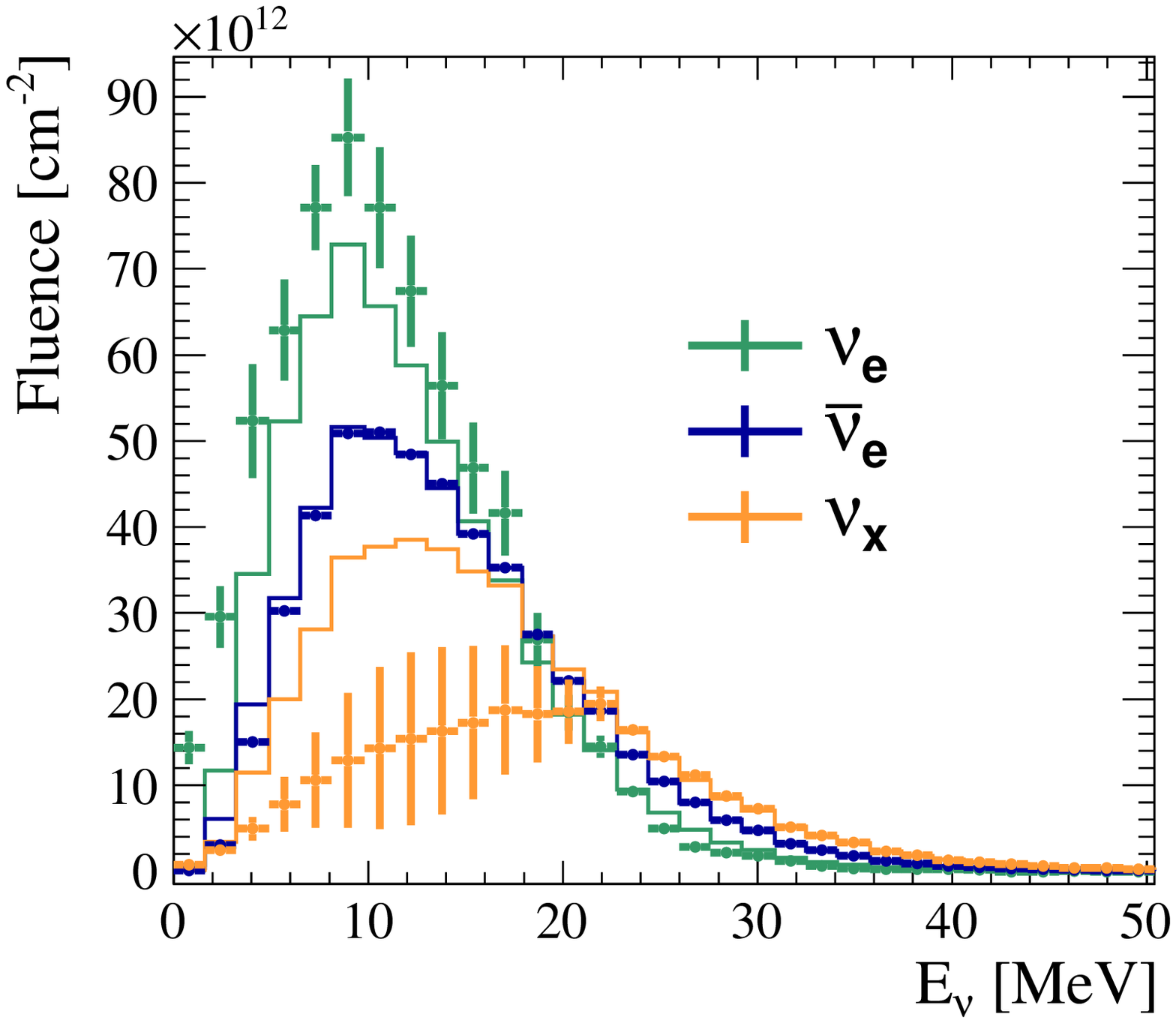}
\end{tabular}
\end{center}
\vspace{-0.5cm}
\caption{The unfolded neutrino spectra from the combined analysis, where the upper, middle and lower rows correspond to the SN distances at 10 kpc, 1 kpc, and 0.2 kpc, respectively. The right column shows the whole neutrino spectra, while the left column focuses on the high-energy parts.
\label{fig:ufdResults}}
\end{figure}
Following the approach of combined analysis in the previous section, we reconstruct the energy spectra of all flavor SN neutrinos by using the simulated events described in Section II. For comparison, the observed spectra are simulated for a SN distance at 10 kpc, 1 kpc, and 0.2 kpc. In addition, we investigate the impact of the energy threshold and the regularization parameter on the reconstruction. To quantitatively assess the model dependence, we also calculate the bias distributions of the reconstructed spectra by repeating the reconstructions for several numerical SN models from the Japan group. In the end, we explain how to reconstruct the original spectra of SN neutrinos in the presence of neutrino flavor conversions.

\subsection{Results for analytical models}

To generate the SN neutrino event spectra, we consider the KRJ parametrization of SN neutrino fluences with $\langle E^{}_{\nu^{}_e}\rangle = 12~{\rm MeV}$, $\langle E^{}_{\overline{\nu}^{}_e} \rangle = 14~{\rm MeV}$ and $\langle E^{}_{\nu^{}_x} \rangle = 16~{\rm MeV}$, and the SN distance is taken to be at 10 kpc, 1 kpc and 0.2 kpc. The observed neutrino events in the IBD, $p$ES and $e$ES channels are then simulated for a JUNO-like detector with $E_{\rm o}^{\rm {th}}=0.2$ MeV. The observed spectra in the $p$ES and $e$ES channels are divided into equal-size bins in a logarithmic scale, while that for IBD is equally binned in a linear scale. The number of bins for IBD, $p$ES and $e$ES is simply taken to be the same, i.e., 20, 30 and 40 bins for the SN at 10 kpc, 1 kpc, and 0.2 kpc, respectively, which can be read out from Fig.~\ref{fig:ufdResults}. Then, the SVD unfolding method is implemented to extract the true neutrino spectra $\bm{F}^{}_{\rm c}$ from the whole observed spectra $\bm{S}^{}_{\rm c}$.

The reconstructed SN neutrino spectra are shown in Fig.~\ref{fig:ufdResults}, where three rows correspond to the case of SNe at 10 kpc (upper panels), 1 kpc (middle panels), and 0.2 kpc (lower panels). The plots in the left column of Fig.~\ref{fig:ufdResults} focus on the high-energy parts (i.e., above about 20 MeV) of the neutrino spectra, while those in the right column represent the full spectra. In all the plots, the solid histograms stand for the true SN neutrino energy spectra while the points denote the reconstructed spectra. The vertical error bars attached to the points indicate the statistical uncertainties, arising from those of the observed SN neutrino spectra, and the horizontal ones show the bin widths. In the cases of the three SN distances, the results are in general better than those obtained in Ref.~\cite{Huiling2018} via a simple bin-by-bin separation method. It is straightforward to understand why the combined analysis works better. First, the cross sections of different flavor neutrinos in $e$ES channel are treated accurately while there is an approximation of $\nu^{}_e$-domination in Ref.~\cite{Huiling2018}. Second, taking account of the correlation among different reaction channels, the combined analysis is able to reduce the meaningless fluctuations in the simple bin-by-bin separation procedure. For the SN at 1 kpc, the best precision for the $\overline{\nu}^{}_{e}$, $\nu^{}_{e}$ and $\nu^{}_{x}$ spectra can reach the level of $1\%$, $20\%$ and $6\%$, respectively. It is obvious that the precision gets better when the SN distance becomes smaller and thus the statistics turns out to be larger.

It is worthwhile to note that the reconstructed $\nu^{}_e$ and $\nu^{}_x$ spectra in the right column of Fig.~\ref{fig:ufdResults} deviates significantly from the true spectra below 20 MeV and they are anti-correlated between each other. The main reason for this behavior is that such low-energy neutrinos can only produce protons with low recoil energies, which after the quenching effects will be lying below the threshold $E_{\rm o}^{\rm{th}} = 0.2~{\rm MeV}$ of the observed energy. Therefore, the $p$ES channel at the LS detector is only sensitive to SN neutrinos with energies above 20 MeV or so. Then it is clear that the deviation is a systematic bias of the energy threshold and the $\nu^{}_{e}$ and $\nu^{}_{x}$ spectra below 20 MeV can be constrained but cannot reconstructed accurately through one single observed spectrum in the $e$ES channel. We shall give a detailed discussion on the effect of different energy thresholds in the next subsection.

\subsubsection{The impact of energy threshold}

\begin{figure}
\begin{center}
\begin{tabular}{c}
\includegraphics[scale=0.7]{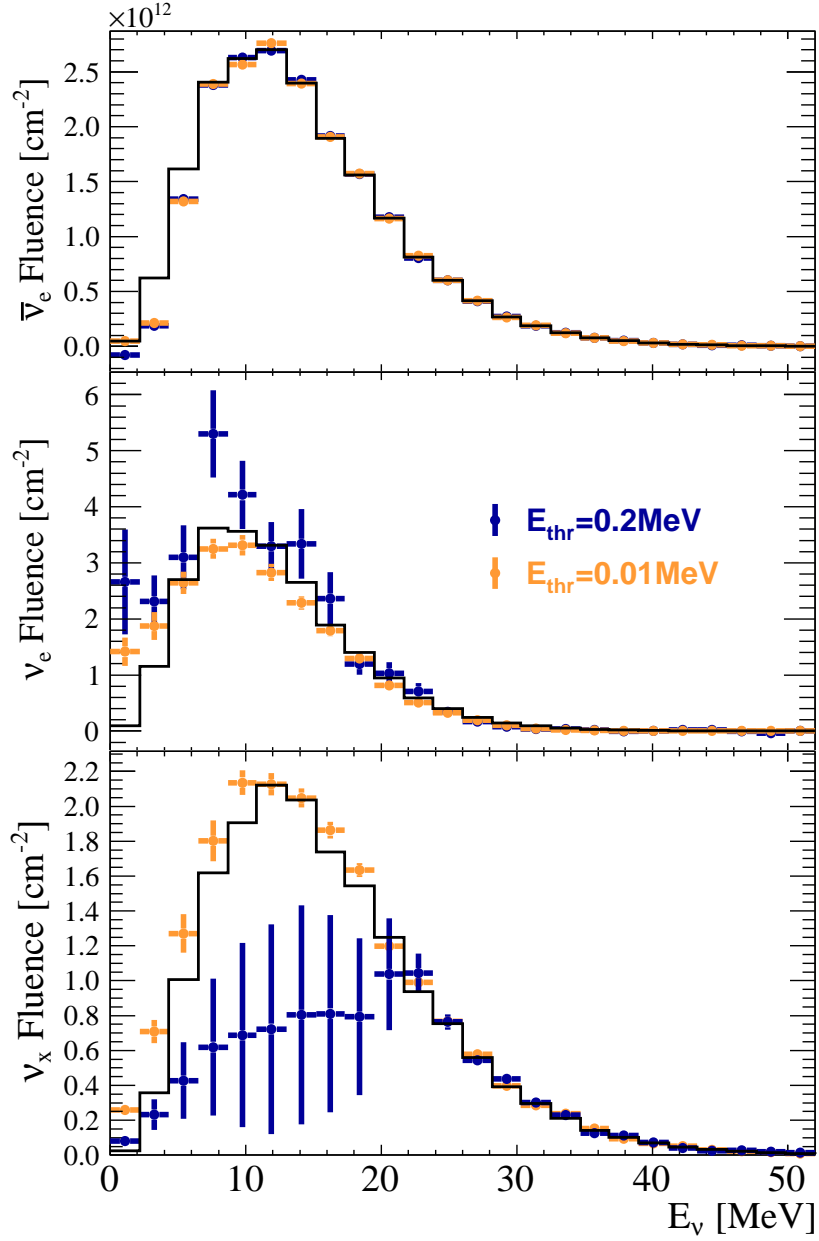}
\end{tabular}
\end{center}
\vspace{-0.7cm}
\caption{The unfolded neutrino spectra from the combined analysis for different flavor neutrinos, where the SN distance is 1 kpc and the energy threshold $E_{\rm o}^{\rm {th}}$ is 0.2 MeV for blue points and 0.01 MeV for orange points.
\label{fig:compThreshold}}
\end{figure}
To check the impact of $E_{\rm o}^{\rm {th}}$ on the reconstruction, we generate another trial of events with $E_{\rm o}^{\rm {th}} = 0.01$ MeV, which is just taken for illustration and will certainly be impossible to achieve in the present and next-generation LS detectors. For the SN at 1 kpc, the unfolded neutrino spectra are shown with orange points in Fig.~\ref{fig:compThreshold}.

For comparison, the corresponding results with an energy threshold of $E^{\rm th}_{\rm o} = 0.2~{\rm MeV}$ have been displayed in the blue points, which is identical to the result in the middle row and right column of Fig.~\ref{fig:ufdResults}. Due to the severe quenching effect of protons in the LS, the minimal energy of incident neutrinos which induce the $p$ES events with the observed energy of 0.01 MeV is about 6.7 MeV, which can also be seen from the spectra of $\nu^{}_{e}$ and $\nu^{}_{x}$ of Fig.~\ref{fig:compThreshold}. The reconstructed spectra of $\nu^{}_{e}$ and $\nu^{}_{x}$ are well consistent with the true spectra for bins above 6.7 MeV. The comparison between these two cases of different values of energy threshold clarifies that a lower threshold of the observed energy can help to extract the $\nu^{}_x$ spectrum accurately in a wider range of energies.

\subsubsection{The impact of regularization parameters}

As usual, the regularization parameter is introduced in the unfolding procedure in order to suppress the spurious oscillatory components which arise from the direct computation of the inverse of the response matrix. In doing so, the bias will be unavoidably brought into the final results at the same time. To optimize the value of the regularization parameter, one should make a balance between the reduction of spurious oscillatory components and the bias. Given the KRJ parameterization, we simulate 500 trials of neutrino events with the SN at 1kpc. Then the combined method is applied to reconstruct the neutrino spectra with a high, medium and low values of the regularization parameter respectively. The bias is defined as the relative difference between the reconstructed and true neutrino spectra in each energy bin, which includes both the statistical fluctuation and the actual reconstruction bias from the unfolding method. In order to reduce the statistical fluctuation, we introduce the mean bias $b_{\rm M}$ over 500 trials and its standard deviation $\sigma_{\rm M}$, namely,
\begin{eqnarray}
b^{i}_{\rm M} &=& \frac{1}{N}\sum^{N}_{k=1} b^{i}_{k}, ~~\sigma^{i}_{\rm M} = \frac{1}{N}\sqrt{\sum^{N}_{k=1}(b^{i}_{k}-b^{i}_{\rm M})^{2}}\; ,\label{eq:meanBias}
\end{eqnarray}
with
\begin{eqnarray}
b^{i}_{k}&=&\frac{n^{\rm true}_{k}(E^{i}_{\nu}) - n^{\rm reconstructed}_{k}(E^{i}_{\nu})}{n^{\rm true}_{k}(E^{i}_{\nu})}\; ,
\label{eq:bk}
\end{eqnarray}
where $k$ is the index for the trial and $N = 500$ is the number of total trials for the KRJ parameterization.
\begin{figure}[!t]
\begin{center}
\begin{tabular}{c}
\includegraphics[scale=0.55]{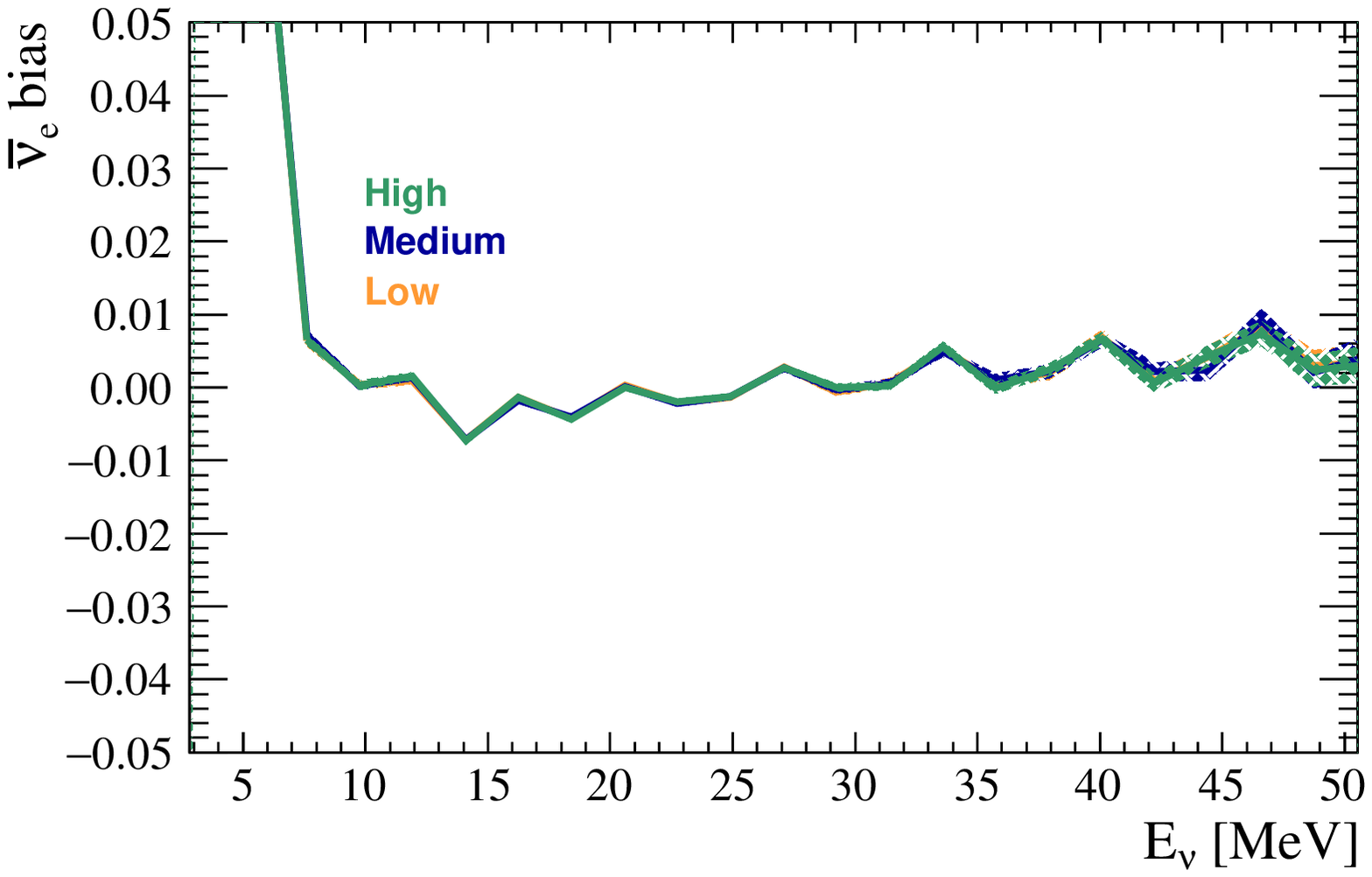}
\\
\includegraphics[scale=0.55]{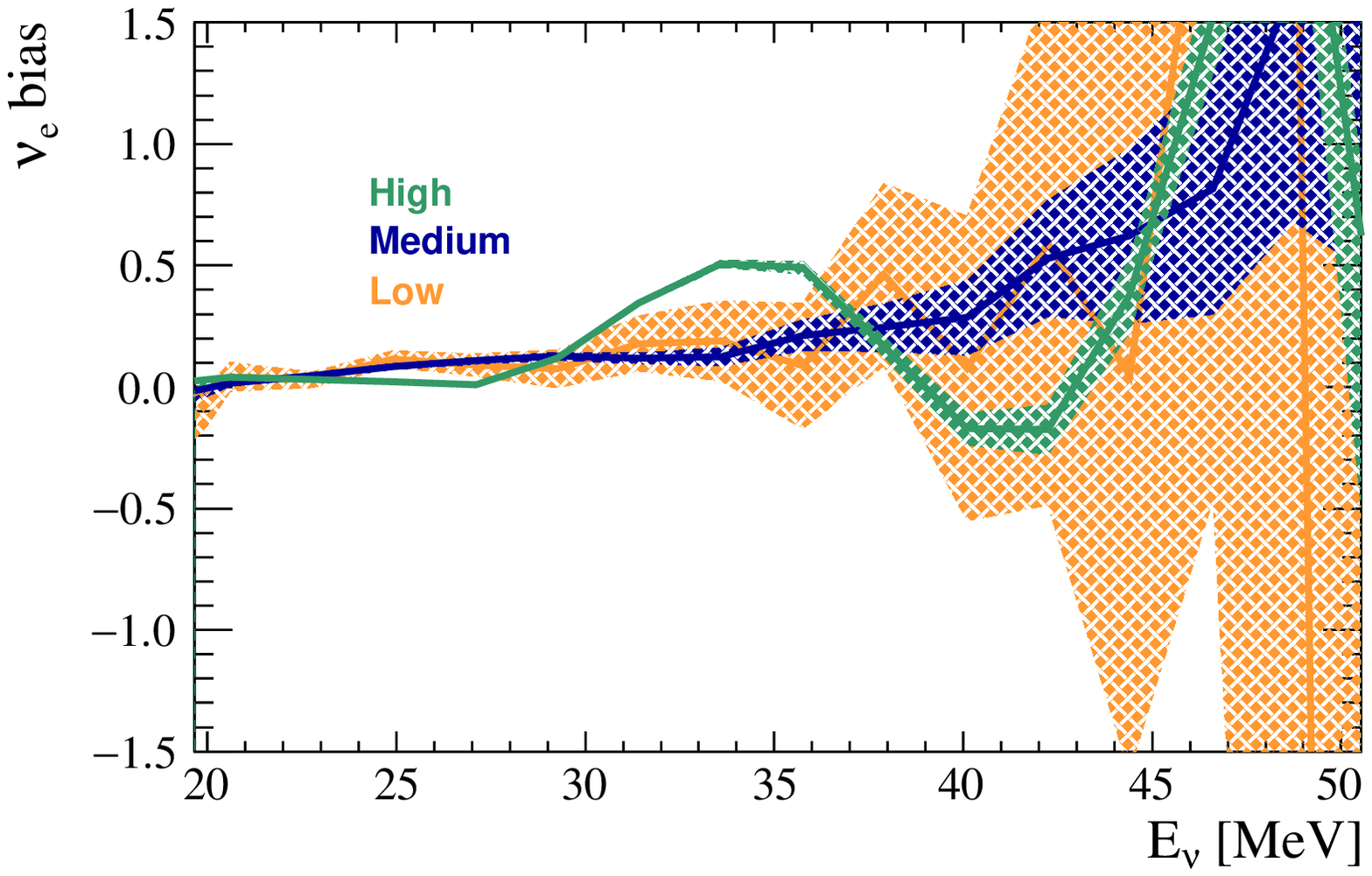}
\\
\includegraphics[scale=0.55]{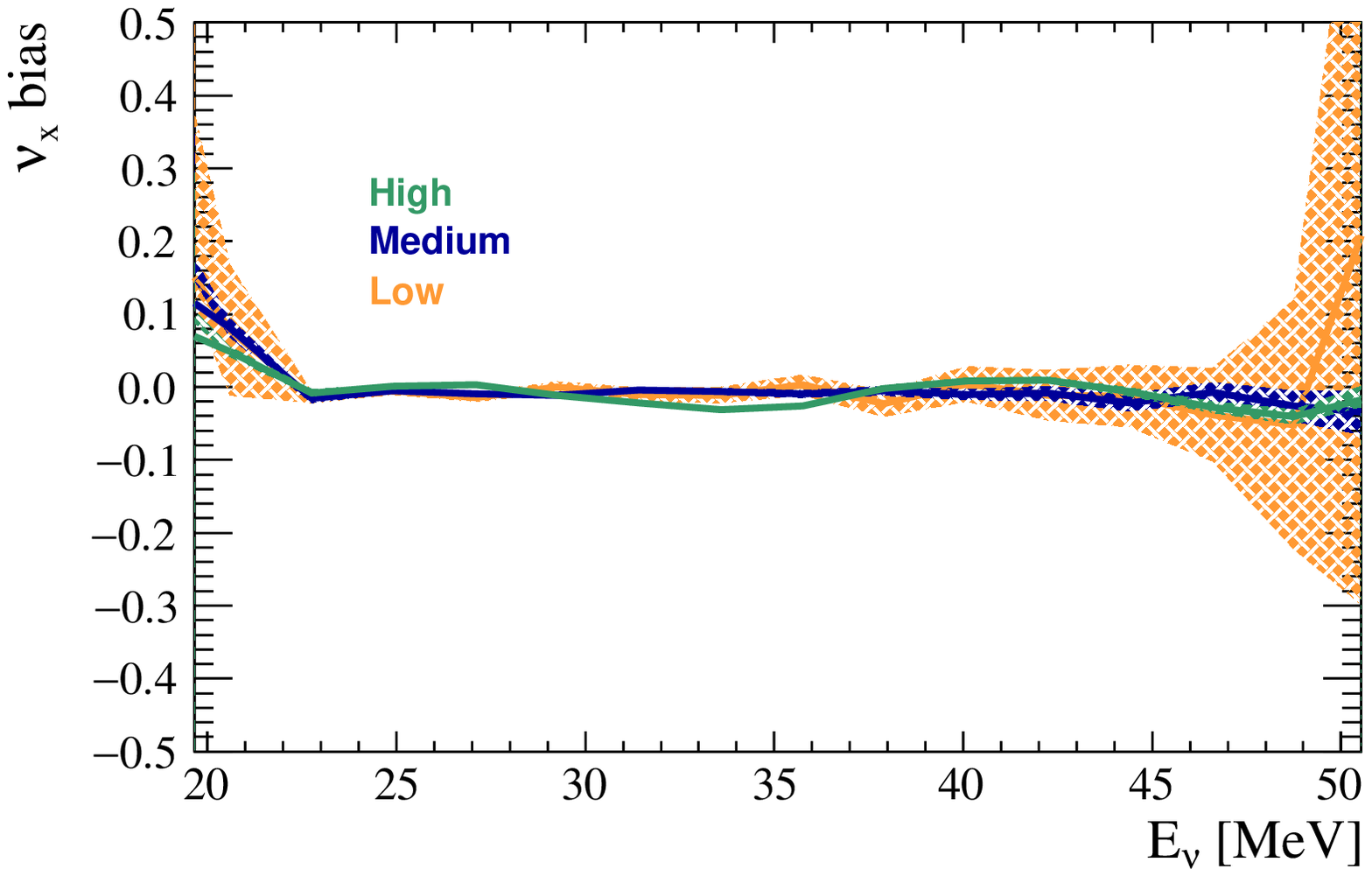}
\end{tabular}
\end{center}
\vspace{-0.6cm}
\caption{The bias distributions for $\overline{\nu}^{}_{e}$, $\nu^{}_{e}$ and $\nu^{}_{x}$ from top to bottom respectively for the SN at 1kpc, where high (green), medium (blue) and low (orange) values of regularization parameter are applied during the unfolding procedure. The solid lines are for the mean bias and the shade areas are due to standard deviations of the mean bias, which are explained in the text.
\label{fig:RegK}}
\end{figure}

In Fig.~\ref{fig:RegK}, the bias distributions for $\overline{\nu}^{}_{e}$, $\nu^{}_{e}$ and $\nu^{}_{x}$ are shown from top to bottom respectively. The solid lines refers to the mean bias distributions $b_{\rm M}$, which can be approximately taken as the bias introduced by the unfolding algorithm. The shade areas are corresponding to the standard deviation of the mean bias $\sigma_{\rm M}$ for the 500 trials with a high (green), medium (blue) and low (orange) values of regularization parameters respectively. The suppression effects are implied from the solid curves and shade areas of different regularization parameters. As the regularization parameter becomes larger, the spurious oscillations in the unfolded spectra are getting more suppressed, but the resultant biases are also more sizable. In the current study, we do not attempt to obtain the optimal value of the regularization parameter for each trial, but directly choose a suitable value to keep the statistical errors of the unfolded spectra smaller than or comparable to the induced bias.

\subsection{Results for numerical models}
\begin{figure}[t!]
\begin{center}
\begin{tabular}{c}
\includegraphics[width=0.5\textwidth]{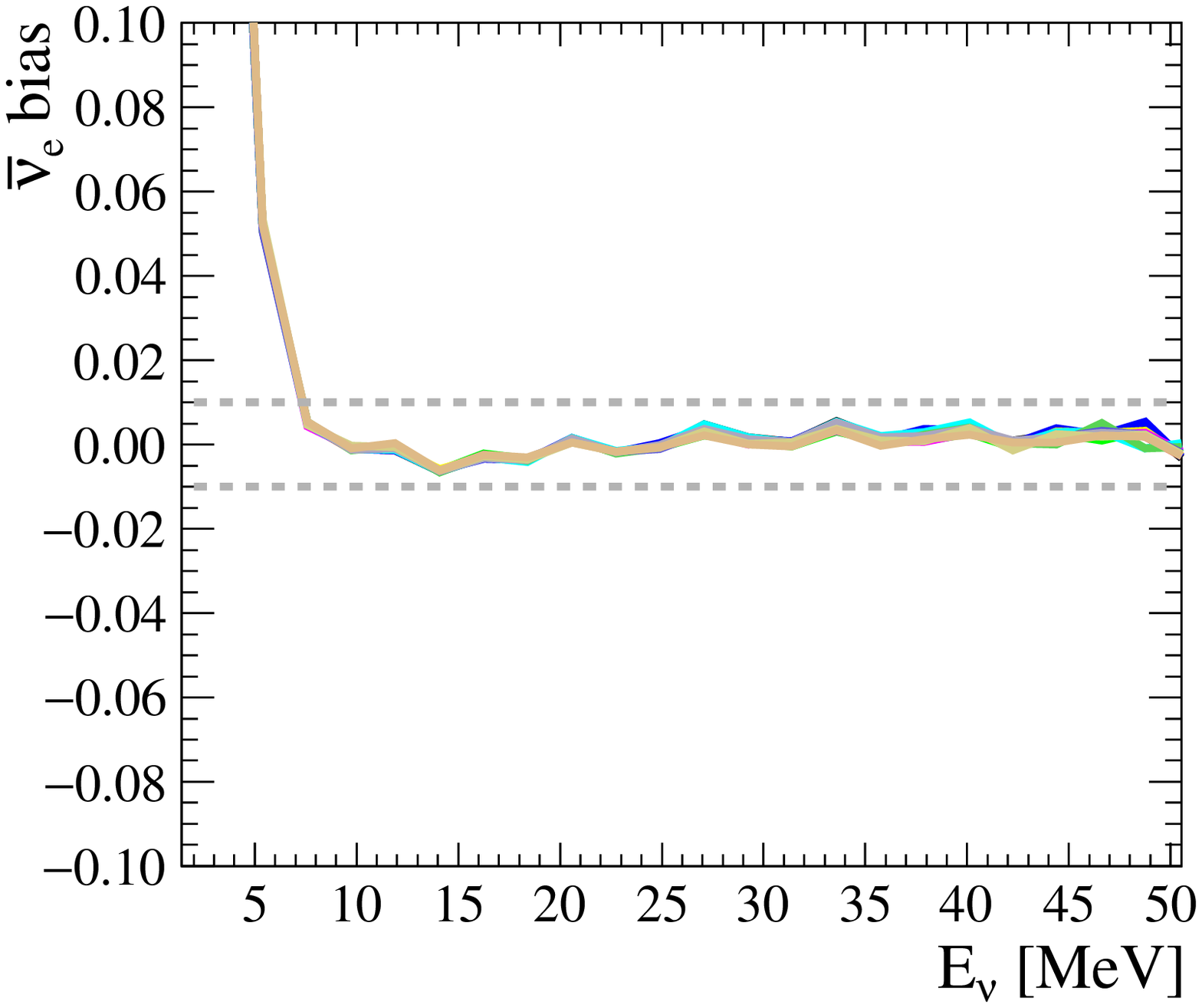}
\includegraphics[width=0.5\textwidth]{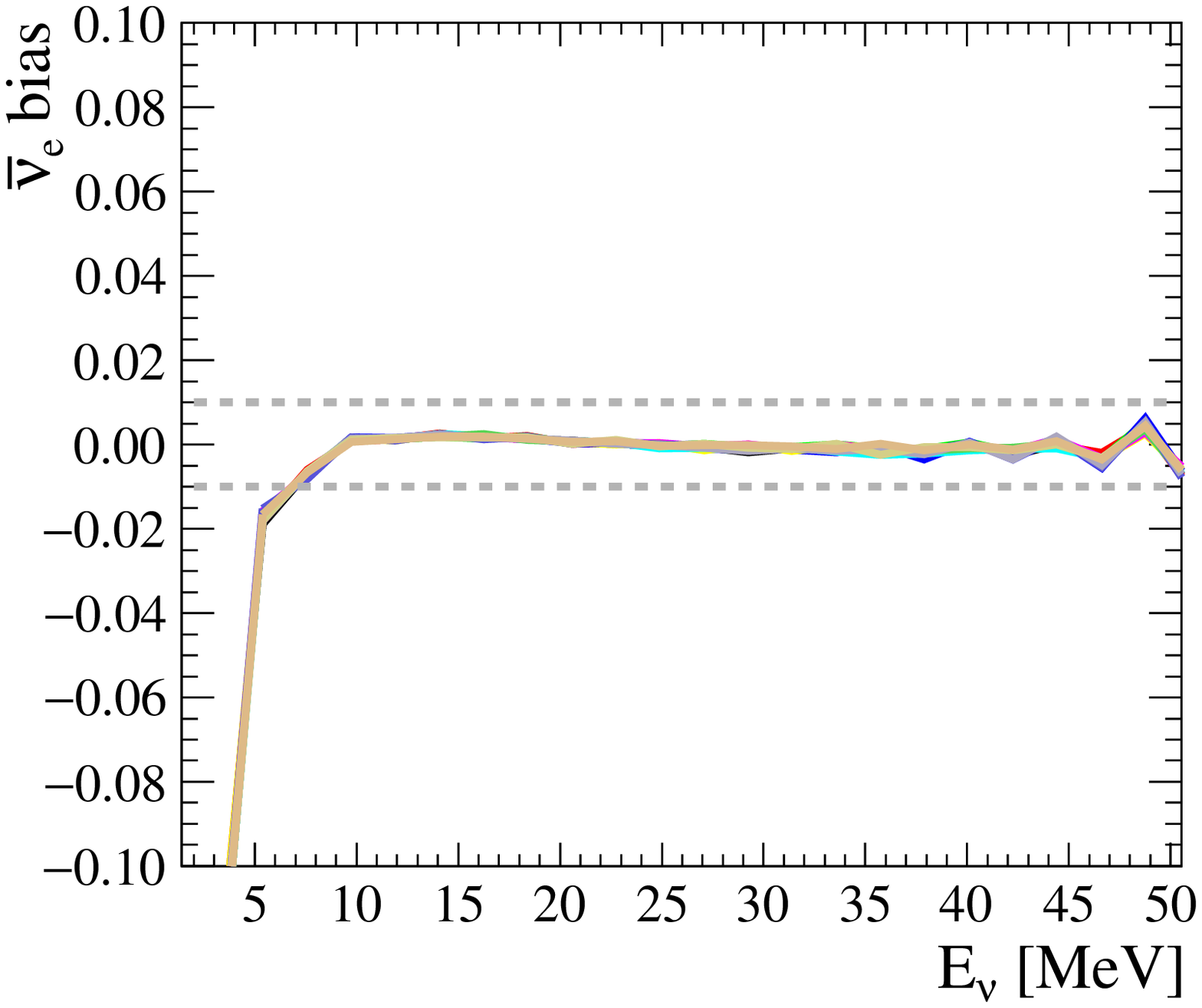}
\\
\includegraphics[width=0.5\textwidth]{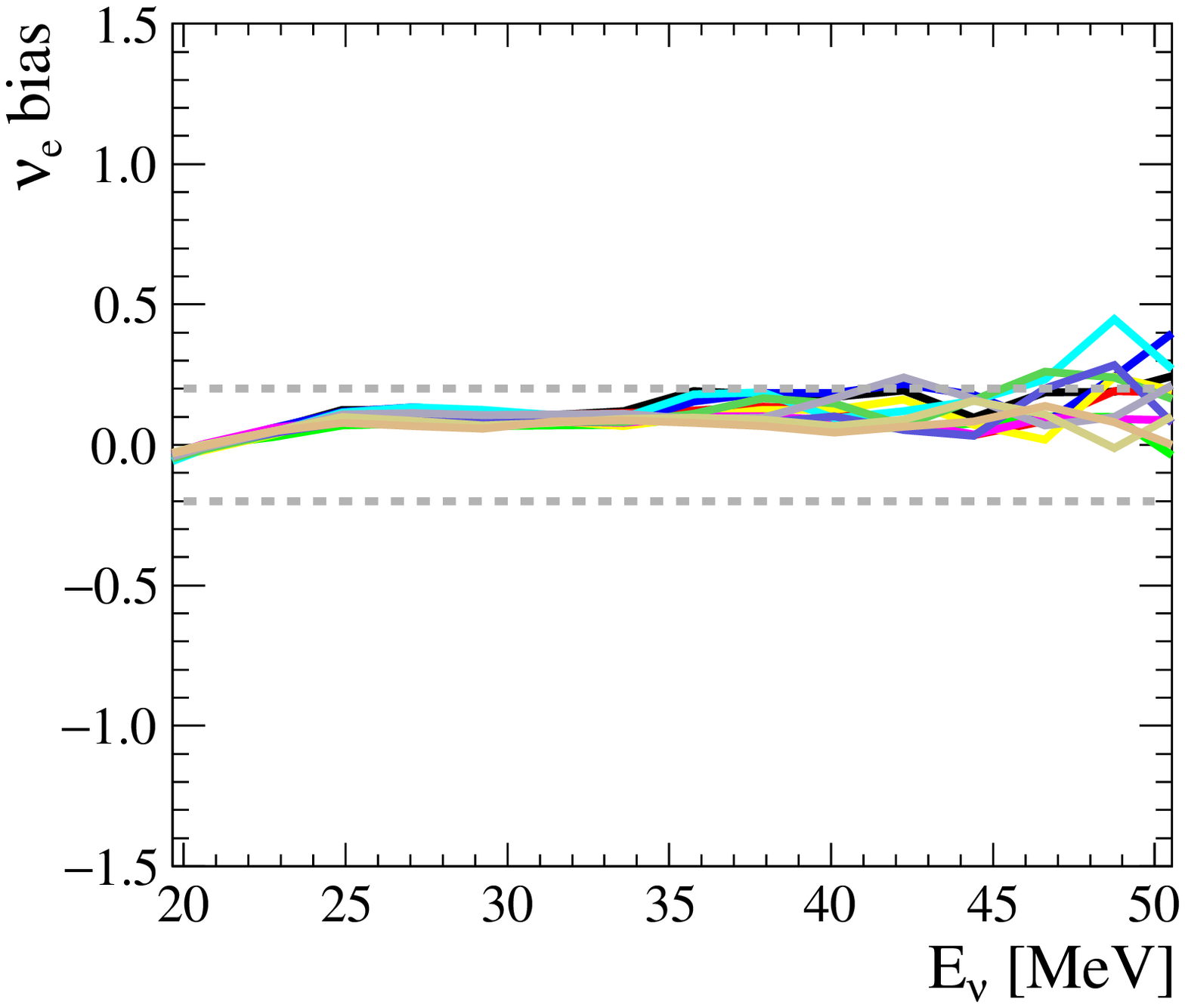}
\includegraphics[width=0.5\textwidth]{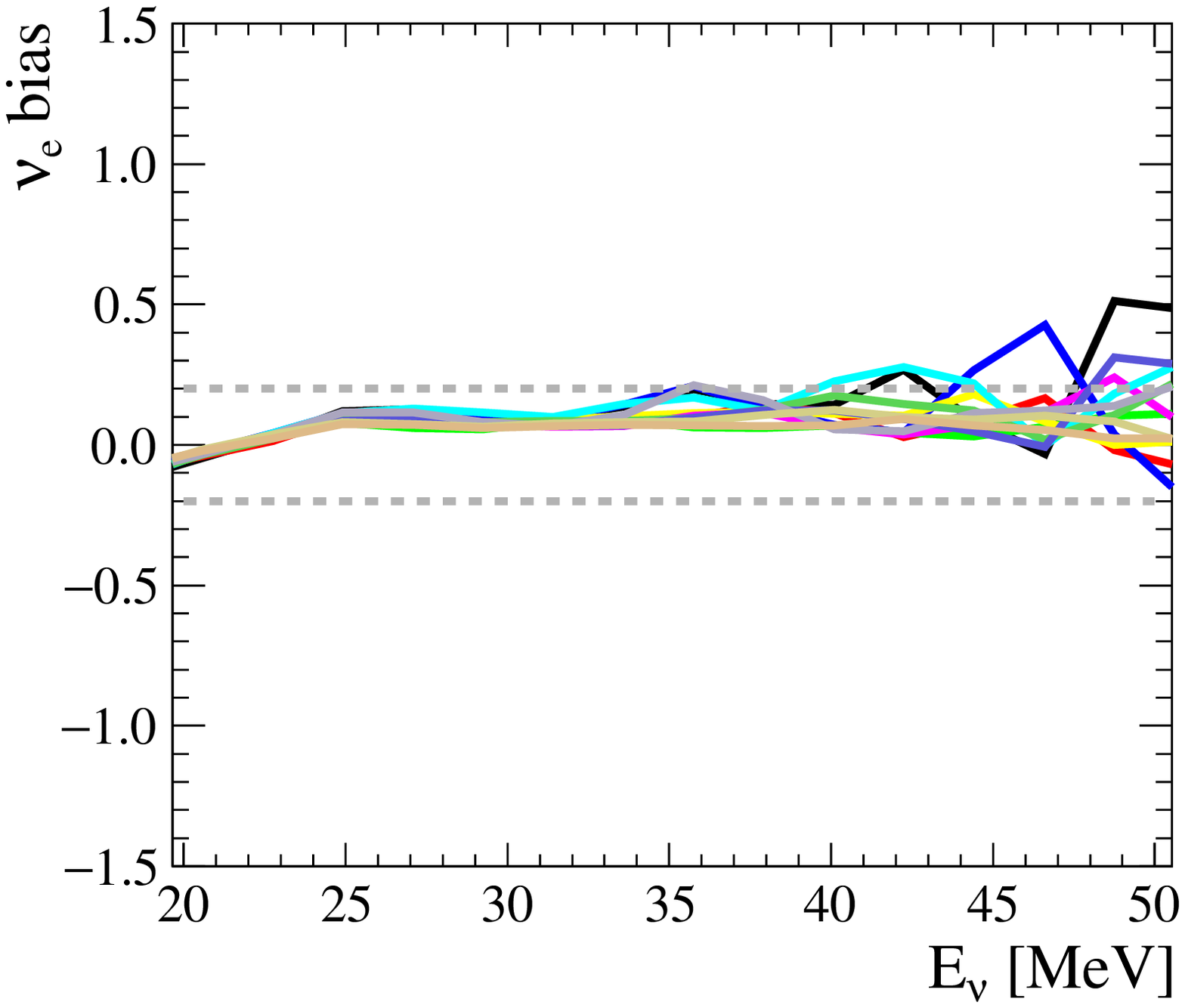}
\\
\includegraphics[width=0.5\textwidth]{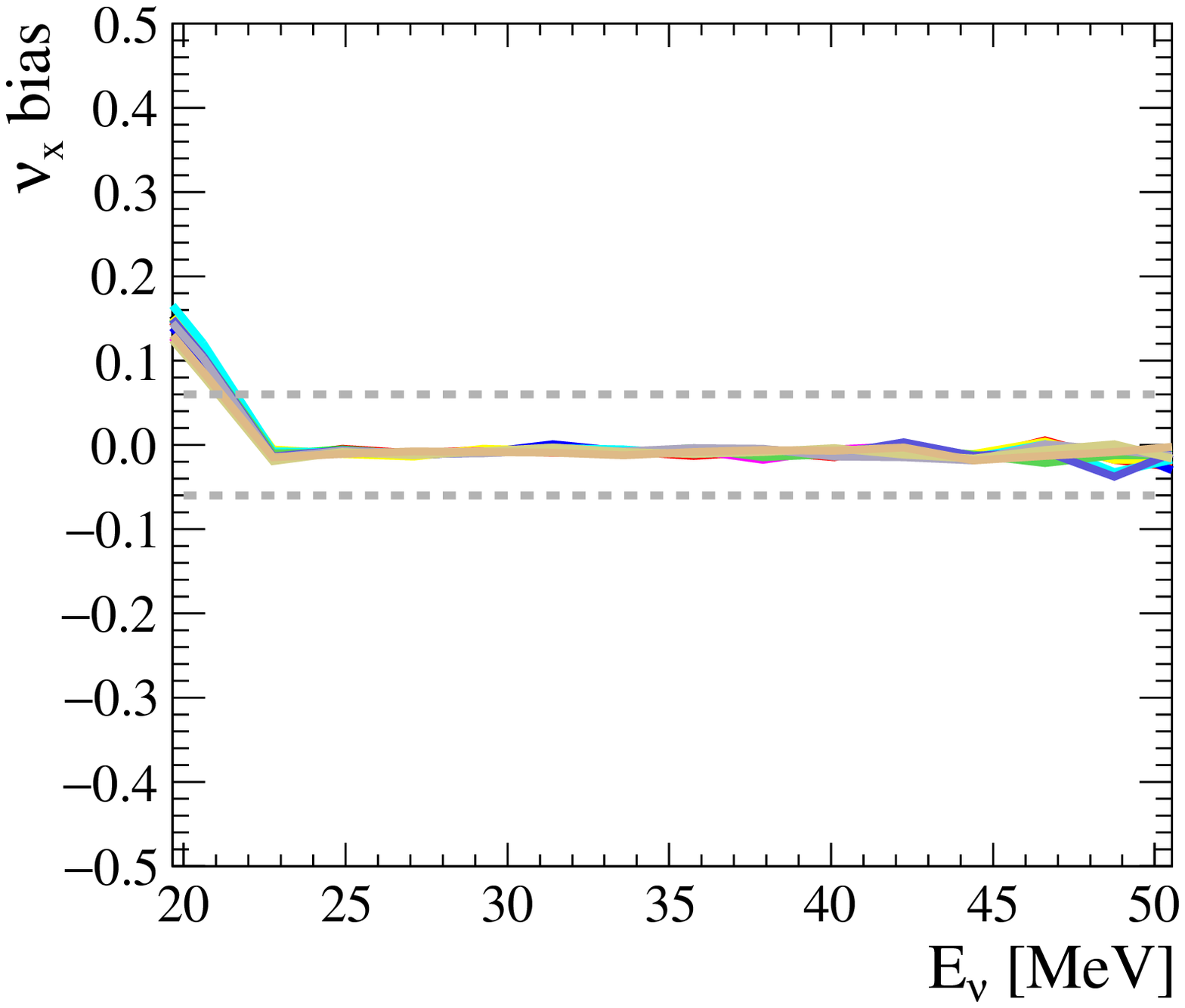}
\includegraphics[width=0.5\textwidth]{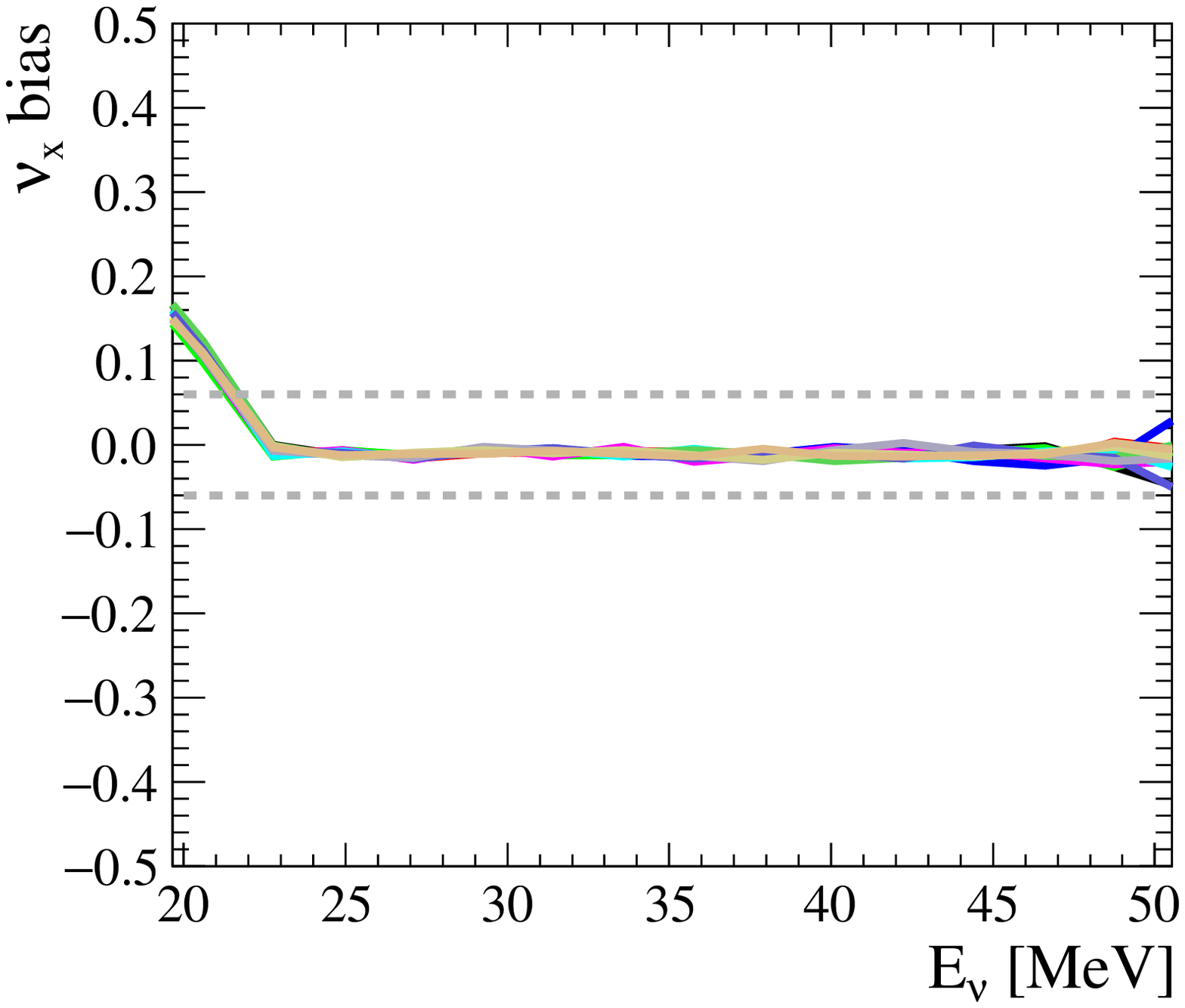}
\end{tabular}
\end{center}
\vspace{-0.8cm}
\caption{The bias distributions of the reconstructed neutrino spectra for the SN at 1 kpc. In the unfolding process, the response matrix has been built with the flat neutrino spectra (left column), or with the neutrino spectra from each numerical model (right column). The dashed horizontal lines are shown for the best precision of the corresponding reconstructed spectra in Fig.~\ref{fig:ufdResults}.
\label{fig:biasJapan}}
\end{figure}

To demonstrate the robustness and the model-independence of the combined method, we implement the SN neutrino fluences from twelve numerical models simulated by the Japan group~\cite{Nakazato:2013}. These models can be classified by the progenitor masses ($M = 13~M^{}_{\odot}$ and 20 $M^{}_{\odot}$), the metallicity ($Z = 0.02$ and 0.004), and the revival time of the shock wave ($t^{}_{\rm riv} = 0.1~{\rm s}$, 0.2 s and 0.3 s). For each numerical model, we simulate 500 trials of neutrino events by assuming the SN at 1 kpc. Then, the combined method is applied to reconstruct the neutrino spectra, with the response matrix built from the flat neutrino energy spectra. In addition, the same calculation is performed again but with a response matrix constructed by using the SN neutrino spectra from this numerical model. The strategy for reconstruction is the same as that adopted for the analytical models. Therefore, for each trial in a specified numerical model, we have the results of unfolded neutrino spectra.

In Fig.~\ref{fig:biasJapan}, the distributions of the mean bias as defined in Eq.~\ref{eq:meanBias} for $\overline{\nu}^{}_{e}$, $\nu^{}_{e}$ and $\nu^{}_{x}$ are shown in the upper, middle and lower row, respectively and the dashed horizontal lines are shown for the best precision of the reconstructed spectra of different flavor neutrinos for the SN at 1 kpc in Fig.~\ref{fig:ufdResults}.  The left column summarizes the results with the fixed response matrix from the flat model, while the right column is for results with the response matrix from the corresponding numerical model. The same value of the regularization parameter is adopted as that for the SN at 1kpc in Fig.~\ref{fig:ufdResults}. Some important conclusions can be drawn. First, the bias distributions for different numerical models, as denoted by the colored curves in Fig.~\ref{fig:biasJapan}, are quite similar to each other, implying that this combined analysis is robust and model-independent. Second, the distributions in left and right column are also well consistent. Although the bias of $\overline{\nu}_{e}$ at a few low-energy bins seems model dependent, the reconstruction for spectra above 20 MeV is robust and not affected by the initial neutrino spectra used to build the response matrix. Third, as seen from those plots in Fig.~\ref{fig:biasJapan}, the averaged bias for the central part of energies in all different flavors is smaller than the statistical errors shown by the dashed horizontal lines. This observation indicates that the adopted regularization parameter is very reasonable.

\subsection{Neutrino flavor conversions}
\begin{figure}[!t]
\begin{center}
\begin{tabular}{l}
\includegraphics[width=0.5\textwidth]{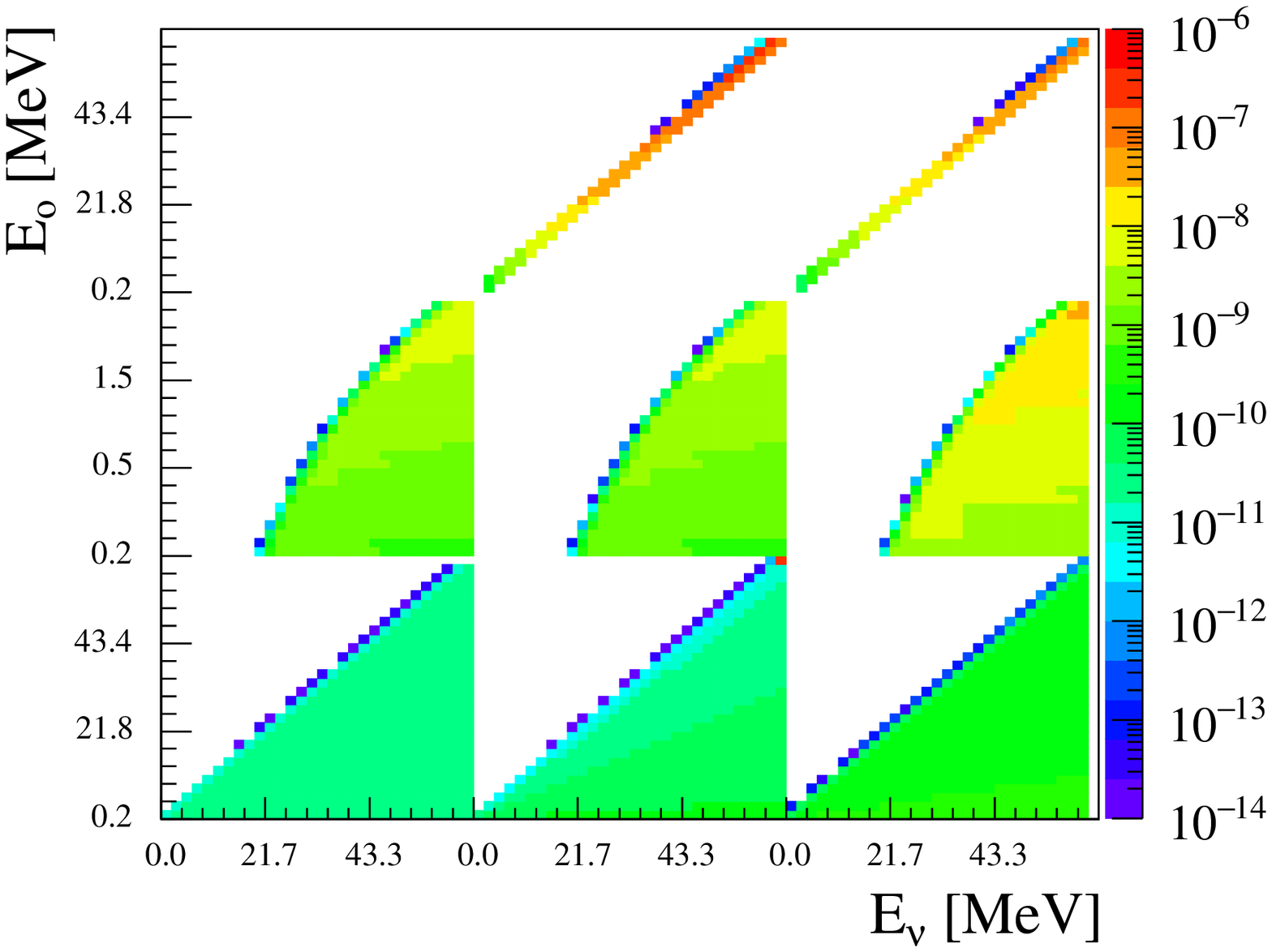}
\hspace{-0.5cm}
\includegraphics[width=0.5\textwidth]{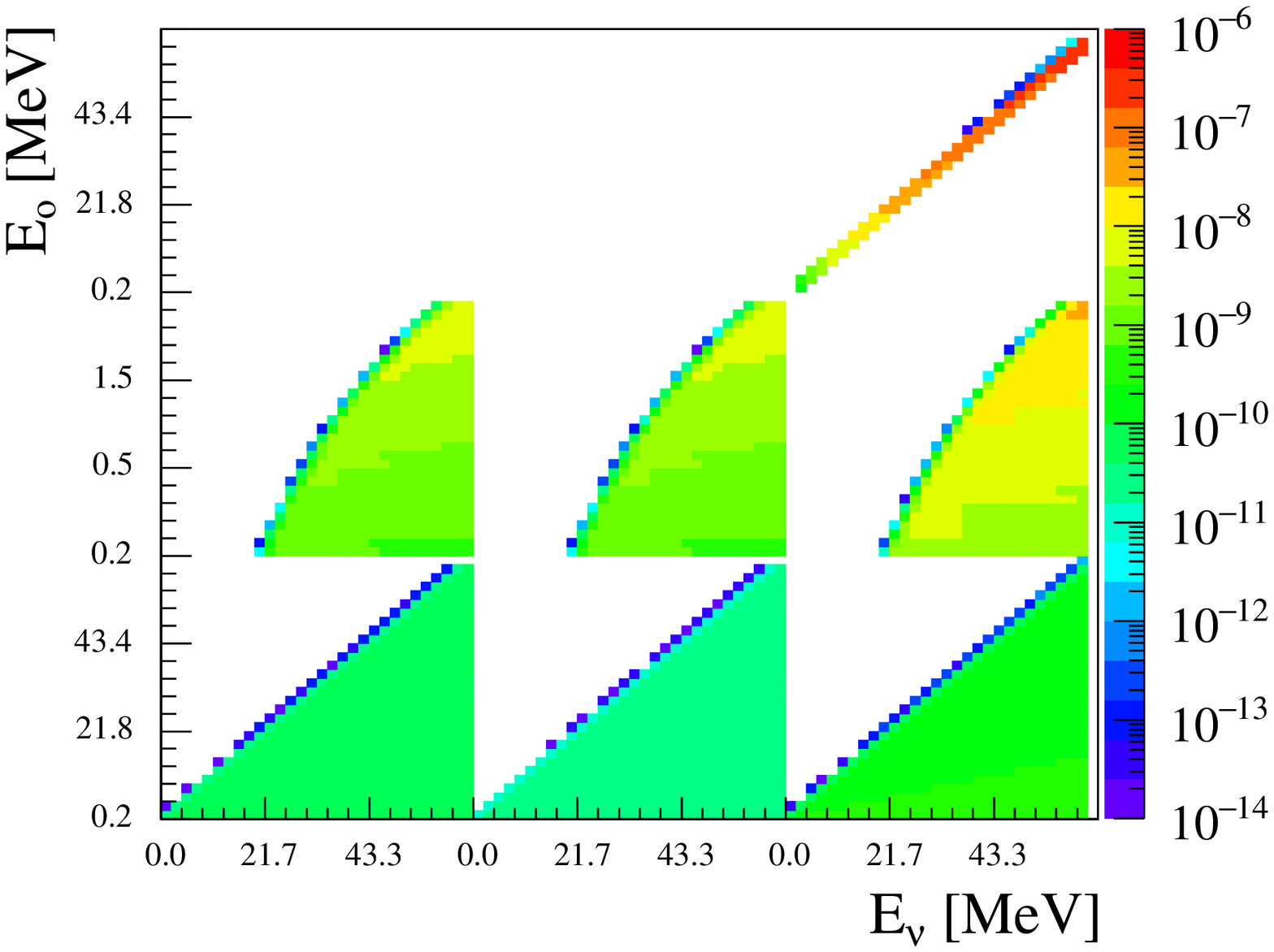}
\end{tabular}
\end{center}
\vspace{-0.5cm}
\caption{The convolution of the response matrices in Fig.~\ref{fig:respMat_noosc} with the flavor conversion matrix in Eq.~(\ref{eq:nhmatrix}) in the case of normal neutrino mass ordering (left panel) or that in Eq.~(\ref{eq:ihmatrix}) in the case of inverted neutrino mass ordering (right panel).
\label{fig:respMatosc}}
\end{figure}
In the previous discussions, the flavor conversions of SN neutrinos have been completely ignored. When propagating outward from the neutrino sphere, SN neutrinos may experience spectral splits or swaps from the collective neutrino oscillations~\cite{Pantaleone:1992eq, Samuel:1993uw, Duan:2005cp, Duan:2006an, Hannestad:2006nj, Raffelt:2007yz, Duan:2009cd, Duan:2010bg, Chakraborty:2016yeg} due to the dense neutrino background. Several recent studies~\cite{Izaguirre:2016gsx, Capozzi201808, Dasgupta2018} show that the self-induced fast oscillations may take place close to the neutrino sphere. However, it remains an open question whether the collective neutrino oscillation do happen in the real SN environment~\cite{Chakraborty:2016yeg}. In the mantle of the SN, the MSW matter effects will play an important role, resulting in a partial or complete conversion between $\overline{\nu}^{}_{e}$ and $\nu_{x}$ (or between $\nu^{}_{e}$ and $\nu^{}_{x}$) depending on the neutrino mass ordering. On the way to the Earth, although SN neutrinos have lost quantum coherence~\cite{Giunti2004, Kersten2016}, there will be regeneration effects due to the Earth matter~\cite{Amol2000, Lunardini:2001pb, Dighe:2003jg, Mirizzi:2006xx, Guo:2006ap, Borriello:2012zc, Liao:2016uis}. Therefore, the final SN neutrino spectrum of a given flavor entering into the detectors is actually a mixture of the initial spectra of different flavors.

To explain how to deal with neutrino flavor conversions in the reconstruction of SN neutrino spectra, we take into account the MSW matter effects~\cite{Amol2000}, for which the overall picture is clear and well understood. Once the collective oscillations of SN neutrinos are established, it will be straightforward to incorporate them into our analysis. According to Ref.~\cite{Amol2000} and the latest neutrino oscillation data~\cite{Capozzi2017}, we can find for the normal neutrino mass ordering (NO)
\begin{equation}
\begin{bmatrix}
F_{\nu^{}_{e}}^{\prime} \\ F_{\overline{\nu}^{}_{e}}^{\prime} \\ F_{\nu^{}_{x}}^{\prime}
\end{bmatrix}
=
\begin{bmatrix}
0 & 0 & 1 \\
0 & \cos^{2}\theta^{}_{12} & \sin^{2}\theta^{}_{12} \\
\displaystyle \frac{1}{4} & \displaystyle \frac{1}{4}\sin^{2}\theta^{}_{12}  & \displaystyle \frac{1}{4}(2+\cos^{2}\theta^{}_{12})
\end{bmatrix} \cdot
\begin{bmatrix}
F^{}_{\nu^{}_{e}} \\ F^{}_{\overline{\nu}^{}_{e}} \\ F^{}_{\nu^{}_{x}}
\end{bmatrix} \; ;
\label{eq:nhmatrix}
\end{equation}
or for the inverted neutrino mass ordering (IO):
\begin{equation}
\begin{bmatrix}
F_{\nu^{}_{e}}^{\prime} \\ F_{\overline{\nu}^{}_{e}}^{\prime} \\ F_{\nu^{}_{x}}^{\prime}
\end{bmatrix}
=
\begin{bmatrix}
\sin^{2}\theta^{}_{12} & 0 & \cos^{2}\theta^{}_{12} \\
0 & 0 & 1 \\
\displaystyle \frac{1}{4}\cos^{2}\theta^{}_{12} & \displaystyle \frac{1}{4}  & \displaystyle \frac{1}{4}(2+\sin^{2}\theta^{}_{12})
\end{bmatrix} \cdot
\begin{bmatrix}
F^{}_{\nu^{}_{e}} \\ F^{}_{\overline{\nu}^{}_{e}} \\ F^{}_{\nu^{}_{x}}
\end{bmatrix} \; ,
\label{eq:ihmatrix}
\end{equation}
where $F^{}_{\alpha}$ stands for the initial neutrino spectra as given by the KRJ parametrization while $F_{\alpha}^{\prime}$ is the spectra after flavor conversions. The conversion matrix between the initial and final spectra will be denoted as $\bm{C}$, in which the neutrino mixing angle $\theta^{}_{12} \approx 33^\circ$ is taken from Ref.~\cite{Capozzi2017}.

\begin{figure}[!t]
\begin{center}
\begin{tabular}{l}
\includegraphics[width=0.55\textwidth]{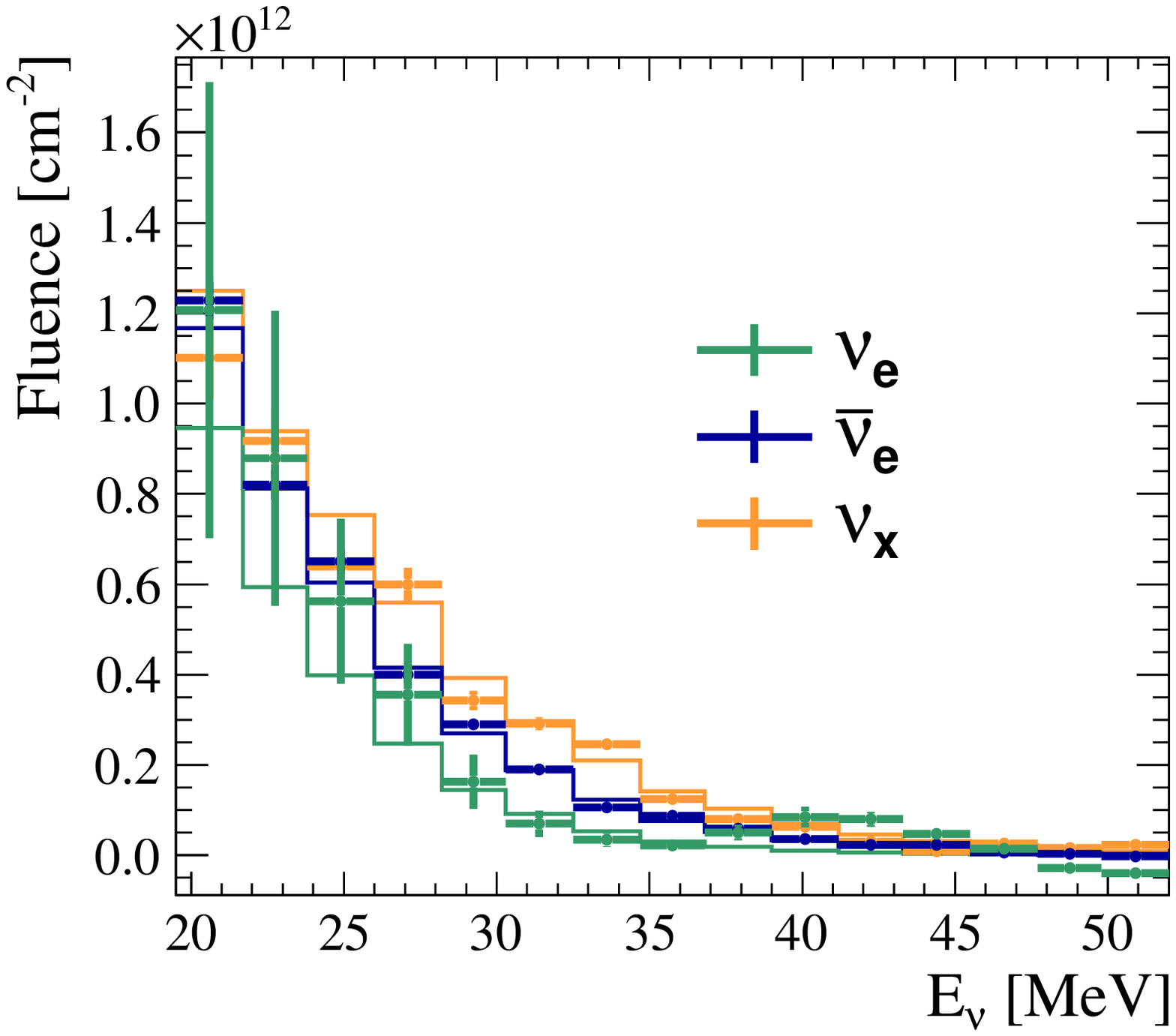}
\hspace{-1.3cm}
\includegraphics[width=0.55\textwidth]{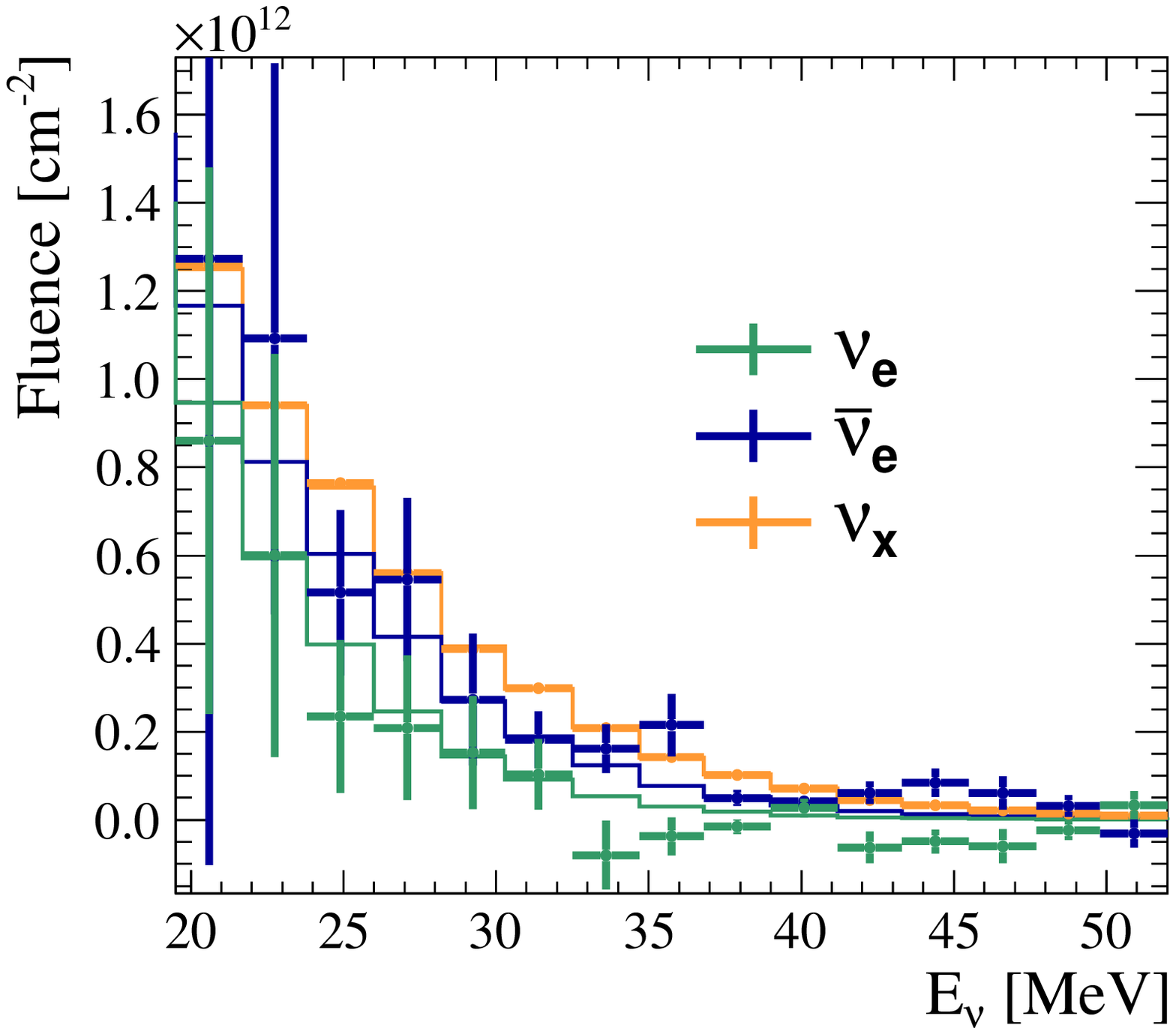}
\end{tabular}
\end{center}
\vspace{-0.5cm}
\caption{The initial SN neutrino spectra reconstructed via the combined analysis, where neutrino flavor conversions are taken into account in the case of the normal mass ordering (left panel) or the inverted mass ordering (right panel).
\label{fig:UfdMH}}
\end{figure}

In consideration of neutrino flavor conversions, the observed event spectra $\bm{S}^{}_{\rm c}$ in the LS detector shall be written as
\begin{equation}
\bm{S}^{}_{\rm c} = \bm{R}^{}_{c} \cdot \bm{F}_{\rm c}^{\prime} \; ,
\end{equation}
where $\bm{F}_{\rm c}^{\prime} = \bm{C} \cdot \bm{F}^{}_{\rm c}$ with $\bm{F}_{\rm c}$ being the initial neutrino spectra. Since the $p$ES channel is subject only to the neutral-current interaction, the event spectrum is not affected by neutrino flavor conversions. However, the event spectra in the IBD and $e$ES channels will differ from those in the scenario without flavor conversions. To extract the initial SN neutrino spectra, we can apply the combined method as well if the response matrix $\bm{R}_{\rm c}$ is convolved with the flavor conversion matrix $\bm{C}$. The results of such a convolution are shown in Fig.~\ref{fig:respMatosc}.

Two trials of SN neutrino events in the IBD, $p$ES and $e$ES channels for a JUNO-like detector are simulated, including neutrino flavor conversions as in Eq.~(\ref{eq:nhmatrix}) and Eq.~(\ref{eq:ihmatrix}) for NO and IO, respectively. The initial neutrino spectra are described by the KRJ parametrization with $\langle E^{}_{\nu^{}_e}\rangle = 12~{\rm MeV}$, $\langle E^{}_{\overline{\nu}^{}_e} \rangle = 14~{\rm MeV}$ and $\langle E^{}_{\nu^{}_x} \rangle = 16~{\rm MeV}$ for a SN distance of 1 kpc. Then, following the combined analysis, we reconstruct the initial SN neutrino spectra and show the final results in Fig.~\ref{fig:UfdMH}. For the IO case, the $\nu^{}_{x}$ spectrum can be well reconstructed because $F^\prime_{\overline{\nu}^{}_e} = F^{}_{\nu^{}_x}$ and $\overline{\nu}^{}_e$ at the detector is precisely measured in the IBD channel. While for the NO case, the IBD channel constrains the spectra of both $\overline{\nu}^{}_e$ and $\nu^{}_{x}$ because $F^\prime_{\overline{\nu}^{}_e} = \cos^2 \theta^{}_{12} F^{}_{\overline{\nu}^{}_e}+\sin^{2} \theta^{}_{12} F^{}_{\nu^{}_{x}}$. Therefore the spectral accuracies of $\overline{\nu}^{}_e$ and $\nu^{}_{x}$ are not as good as that of $\nu^{}_{x}$ in the IO case.
When a more complicated scenario of neutrino flavor conversions is considered, one can replace the flavor conversion matrix $\bm{C}$ with the new one and repeat the analysis to extract the initial SN neutrino spectra. Therefore, our strategy for reconstruction is useful to test the pattern of SN neutrino flavor conversion. Note that the current study of neutrino oscillation effects is only included in the time-integrated neutrino spectra reconstruction. The scenario of neutrino flavor conversions might be different for different phases of the SN neutrino burst. We want to stress that the method proposed here can still be applicable for the reconstruction of the time-dependent neutrino energy spectra, but one needs to suffer from the relatively lower event statistics.

\section{SUMMARY}
For a future galactic core-collapse SN, we have proposed a model-independent approach to reconstruct all flavor SN neutrino energy spectra by performing a combined analysis of the IBD, $p$ES and $e$ES detection channels in a 20 kiloton JUNO-like LS detector. First of all, we briefly recap the calculation of SN neutrino events in the LS detector and the separated method used in Ref.~\cite{Huiling2018} to reconstruct SN neutrino spectra, where however the response matrix for different neutrino flavors in $e$ES channel is not fully considered. Then, the combined method is introduced to treat all neutrino flavors $\nu^{}_{e}$, $\overline{\nu}^{}_{e}$ and $\nu^{}_{x}$ in three channels on the same footing, and applied in the spectral reconstruction with the simulated SN neutrino events. Similar calculations have been carried out for the SN at different distances (i.e., 10 kpc, 1 kpc, and 0.2 kpc). In addition, we investigate the impact of the threshold energy of the detector and the regularization parameter of the unfolding method on the spectral reconstruction. The combined method is demonstrated to be robust and model-independent via the analysis of both analytical and numerical neutrino data. Finally, taking account of neutrino flavor conversions under the MSW matter effects in the SN mantle, we explain how to implement the combined analysis to extract the initial neutrino spectra in the presence of flavor conversions.

Although we have concentrated on the spectral reconstruction with a LS detector, the detections from the large WC and LAr-TPC detectors should be utilized to probe the SN neutrino spectra globally~\cite{Capozzi201806}. It is intuitively convenient for the combined method to accomplish this task. What one has to do is just to extend the the response matrix and the observed event spectra with the information from other detectors and the relevant detection channels. In this case, the low statistics in the $e$ES channel of the LS detectors can be compensated by the WC and LAr-TPC detectors. On the other hand, the advantage of the low energy threshold of LS detectors is maintained to reconstruct the $\nu^{}_{x}$ spectrum. The combined analysis of the LS, WC and LAr-TPC detectors in the reconstruction of SN neutrino spectra is very interesting and deserves another dedicated study. Moreover, this method can also be used to reconstruct the spectra of solar neutrinos and ultrahigh-energy cosmic neutrinos when a large statistics in the multi-flavor detection is accumulated.

\section*{ACKNOWLEDGEMENTS}
This work was supported in part by the National Key R$\&$D Program of China under Grant No. 2018YFA0404100, by the Strategic Priority Research Program of the Chinese Academy of Sciences under Grant No. XDA10010100,  by the National Natural Science Foundation of China under Grant No. 11775232 and No. 11835013, and by the CAS Center for Excellence in Particle Physics (CCEPP).

\end{document}